\documentclass[%
 reprint,
 superscriptaddress,
 amsmath,amssymb,
prb,
]{revtex4-2}
\usepackage{todonotes}
\usepackage{soul}

\usepackage{hyperref}

\usepackage{graphicx}

\def\presuper#1#2%
{\mathop{}%
	\mathopen{\vphantom{#2}}^{#1}%
	\kern-\scriptspace%
#2}

\usepackage{wrapfig}
\usepackage{tabularx}
\usepackage{capt-of}
\usepackage{color}
\usepackage{subfigure}
\usepackage{multirow}
\usepackage{array}
\graphicspath{{images/}}
\usepackage{amsmath,amssymb,amsfonts,textcomp}
\usepackage{mathtools}
\usepackage{physics}

\usepackage{color}
\usepackage{subfigure}
\usepackage{multirow}
\usepackage{array}
\usepackage{outlines}    

\usepackage{algpseudocode}
\usepackage{algorithm2e}

\usepackage{float}
\setcitestyle{super}

\newcommand{\TTq}{\prescript{5}{}{\left(\mathrm{TT}\right)} }
\newcommand{\TTt}{\prescript{3}{}{\left(\mathrm{TT}\right)} }
\newcommand{\TTs}{\prescript{1}{}{\left(\mathrm{TT}\right)} }

\begin{document}

\title{Generalization of the tensor product selected CI method for molecular excited states}

\author{Nicole M. Braunscheidel}\thanks{These two authors contributed equally}
\affiliation{Department of Chemistry, Virginia Tech,
Blacksburg, VA 24060, USA}
\author{Vibin Abraham}\thanks{These two authors contributed equally}
\affiliation{Department of Chemistry, University of Michigan, Ann Arbor, MI 48109, USA}

\author{Nicholas J. Mayhall}
\email{nmayhall@vt.edu}
\affiliation{Department of Chemistry, Virginia Tech,
Blacksburg, VA 24060, USA}

\begin{abstract}
\noindent\textbf{Abstract:} 
In a recent paper [\textit{JCTC}, \textbf{2020}, \textit{16}, 6098],
we introduced a new approach for accurately approximating full CI ground states in large electronic active-spaces,
called Tensor Product Selected CI (TPSCI). 
In TPSCI, a large orbital active space is first partitioned into disjoint sets (clusters) for which the exact, local many-body eigenstates are obtained. 
Tensor products of these locally correlated many-body states are taken as the basis for the full, global Hilbert space. 
By folding correlation into the basis states themselves, the low-energy eigenstates become increasingly sparse, creating a more compact selected CI expansion.
While we demonstrated that this approach can improve accuracy for a variety of systems,
there is even greater potential for applications to excited states, particularly those which have some excited state character. 
In this paper, we report on the accuracy of TPSCI for excited states, including a far more efficient implementation in the Julia programming language.
In traditional SCI methods that use a Slater determinant basis, accurate excitation energies are obtained only after a linear extrapolation and at a large computational cost.
We find that TPSCI with perturbative corrections provides accurate excitation energies for several excited states of various polycyclic aromatic hydrocarbons (PAH) with respect to the extrapolated result (i.e. near exact result).
Further, we use TPSCI to report highly accurate estimates of the lowest 31 eigenstates for a tetracene tetramer system with an active space of 40 electrons in 40 orbitals,
giving direct access to the initial bright states and the resulting 18 doubly excited (biexcitonic) states.
\end{abstract}

\maketitle

\section{Introduction}

Electronic excited states play an important role in a vast number of technologically relevant processes ranging from solar cells, to sensing, to artificial photosynthesis, and beyond.
Theoretical simulations are key for the interpretation and prediction of spectra, lending detailed support to experiments.
However, not all excited states are easily simulated computationally. 
Traditional theoretical methods that depend on single excitations (common to all linear-response methods) like time-dependent density functional theory (TDDFT)\cite{tddft1984,Marques2004,Furche2002} often fail to properly describe
 charge transfer (CT) states\cite{Dreuw2004,Levine2006,Peach2008} and require an additional doubly excited component to capture the presence of doubly excited states\cite{maitra2021double,Neepa2004,ELLIOTT2011}.
Even more sophisticated  methods like equation of motion coupled cluster with singles and doubles (EOM-CCSD) \cite{geertsen1989,comeau1993} can fail for doubly excited states with errors around 1 eV,\cite{watts1996coupled,Loos2019}
requiring higher excitations to produce accurate results. 
In order to provide qualitatively correct descriptions of two-electron excitations, multireference methods, such as complete active space self-consistent field (CASSCF),\cite{ROOS1980} complete active space second-order perturbation theory (CASPT2),\cite{Andersson1990,FINLEY1998} or multireference configuration interaction (MRCI)\cite{Siegbahn1983,Werner1982} are required.
However, these methods cannot be used for active spaces larger than about 20 orbitals with 20 electrons.
It is also very difficult to select active orbitals for state averaging when
	the ground and excited states differ significantly in dipole moment, seen usually in cases with charge transfer excitations.

Selected configuration interaction (SCI)\cite{Huron1973} based approaches have been recently
	used to calculate accurate estimates for vertical excitation energies\cite{Holmes2017,Loos2018,Chien2018,loos2020medium}, double excitations\cite{Loos2019}, doublet-doublet transitions in radicals\cite{loos2020mountaineering}, and excited state dipole moments and oscillator strengths\cite{loos2023}.
Motivated by the fact that low energy eigenstates often have most of their weight on a relatively small subspace of determinants, SCI techniques 
attempt a bottom-up discovery of this space of ``important'' Slater determinants.
For weakly correlated systems, SCI provides an incredibly efficient approach for obtaining near-FCI estimates of energies and excitation energies. 
However, the computational cost of SCI approaches are heavily dependent on the amount of correlation present, as this necessarily increases the dimension of the important subspace of determinants. 
Although these cited applications have been on small to medium molecules, the SCI variational spaces for these systems are already in tens of millions.
For larger systems, the problem will quickly become intractable for SCI based approaches.

Fortunately, the dimension of the important variational space is not an intrinsic characteristic of a given Hamiltonian, 
but is rather a basis dependent quantity. 
For a trivial example, consider the case where one first rotates the basis into the exact eigenbasis. 
In this basis, the relevant variational space has dimension equal to one.
As such, it is possible to decrease the size of the variational space by ``simply'' choosing a more appropriate basis in which to represent the problem. 
With orbital rotations being the simplest change of basis possible where the many-electron transformation is parameterized by simple one-electron functions, SCI calculations are often performed using the natural orbitals computed from either a cheaper SCI calculation or other single reference methods like CCSD or MP2. 
Even though this does generally lead to a smaller variational dimension compared to using canonical Hartree-Fock orbitals, the improvements are often rather limited to a factor of 2 or so.\cite{Zhang2020,Levine2020} 
Recently, orbital optimization has also been proposed to improve the SCI energies with respect to the number of determinants in a given orbital space.\cite{Levine2020,Smith2017,Yao2021a, loos_oo_cipsi}

Along this direction, we recently introduced a new method called Tensor Product Selected Configuration Interaction (TPSCI)\cite{Abraham2020}
which defines a SCI algorithm not in a Slater determinant basis, but rather, in a basis of tensor product states of locally entangled many-body wavefunctions. 
This amounts to a change of basis, where \textit{many-body rotations} are applied locally to the basis of Slater determinants, folding in local electron correlation into the basis functions themselves. 

 Traditional Selected CI methods are memory limited due to the size of the variational space needed to reach a target accuracy. In TPSCI, our goal is to trade off some run time (TPSCI calculations are significantly slower than Slater determinant methods) for reduced memory requirements (TPSCI variational spaces are generally much smaller than Slater determinant methods).

Other methods such as active space decomposition (ASD)\cite{parkerCommunicationActiveSpace2014,parkerCommunicationActivespaceDecomposition2013} and rank-1 matrix product states \cite{Soichiro2019, vpscasscf} also have a similar framework, operating in a similar tensor product space. 
In ASD, the rapid growth of the Hilbert space was controlled with a low-rank matrix product state (MPS) approximation, instead of a sparsity-based approximation used in our current work. 
While a MPS approximation can be effective for compressing a state, this does impose an often artificial one-dimensional entanglement structure. 
In the rank-1 matrix product state method, the global states are written as a linear combination of entangled states, similar to TPSCI, but mainly focus on disjoint molecular units.
A broad list of methods exist which focus on forming the wavefunction of the full system in this clustered framework, including: 
Block Correlated Coupled Cluster (BCCC)\cite{liBlockcorrelatedCoupledCluster2004}
and the related Tensor Product State Coupled Electron Pair-Type Approximations (TPS-CEPA),\cite{abrahamCoupledElectronPairType2022}
the cMF-based coupled cluster,\cite{Scuseria2022} 
the ab inito Frenkel-Davydov model \cite{Morrison2014,Morrison2015}, 
renormalized exciton model (REM) \cite{AlHajj2005a,Ma2012}, 
Block Interaction Product State (BIPS)\cite{wang2021low}, 
comb-Tensor network states based approach by Li\cite{li2021expressibility}, 
generalized and localized active space methods\cite{ma2011generalized,weser2021stochastic,hermes2020variational}.

In this work, we extend our recently proposed TPSCI methodology\cite{Abraham2020} to provide near-FCI approximations to relatively large manifolds of excited states in a limited basis of active orbitals. 

\section{Methods}
The core strategy in TPSCI is to build a localized representation that increases the sparsity of the target global eigenstates.
Let us start by assuming that our orbital active space permits a  partitioning  into smaller, disjoint  active spaces (referred to as ``clusters'' throughout). 
While clusters can be defined through different considerations (locality, orbital entanglement~\cite{Barcza2011}, symmetry etc.), the general guideline is that the intra-cluster interactions should be stronger than inter-cluster interactions.

Within each cluster, we want to define a many-body transformation\footnote{This transformation can be unitary if we retain the size of the local Hilbert space by not discarding any degrees of freedom. However, more generally, and for larger clusters, this transformation will discard high energy states, making it an isometry mapping a large space to a smaller space, retaining lengths.} that accounts for all relevant local correlations.
In principle, one can obtain such a transformation by  simply diagonalizing the local Hamiltonian (the terms that remain after removing operators that act outside of the cluster). 
However, this explicitly neglects the influence of neighboring clusters on the composition of our many-body transformations. 
We instead include the influence of inter-cluster interactions in a mean-field fashion by adopting the cluster Mean Field (cMF) method that was introduced by Scuseria and coworkers\cite{Scuseria2015, Scuseria2022} and  explored by Gagliardi and coworkers.\cite{Gagliardi2020, pandharkarLocalizedActiveSpaceState2022}

This mean-field treatment arises (analogously to Hartree-Fock theory) by variationally minimizing the energy of a single tensor product state (TPS) 
with respect to both orbital and local many-body rotations (defined by a set of local configuration interaction coefficients).
As such, cMF can be understood as a CASSCF problem  with multiple active spaces,  
similar to generalized active space or occupation restricted active space methods.\cite{ma2011generalized,weser2021stochastic,Ivanic2003}
We will express the cMF ground state wavefunction as:
\begin{equation}
\ket{\psi_{0}} =   \ket{0_{I}}\ket{0_{J}}\dots\ket{0_{N}} = \ket{0_{I}0_{J}\dots 0_{N}}
\end{equation}
where $I$, $J$, $\dots$ label clusters, and $\ket{0_I}$ is the lowest energy eigenstate of the cMF effective Hamiltonian on cluster $I$:
\begin{align}\label{eq:Heff}
    \hat{H}^{eff}_I = & \sum_{pq}^{\in I}h_{pq}\hat{p}^\dagger\hat{q} + \tfrac{1}{2}\sum_{pqrs}^{\in I}\braket{pq}{rs}\hat{p}^\dagger\hat{q}^\dagger\hat{s}\hat{r}\nonumber\\
    & + \sum_{pr}^{\in I}\sum_{J\neq I}\sum_{qs}^{\in J}\expval{pq||rs}\gamma_{qs}^J\hat{p}^\dagger\hat{r},
\end{align}
where $\hat{p}$ ($\hat{p}^\dagger$) are the fermionic annihilation (creation) operators on orbital $p$, $\gamma_{qs}^J$ is an element of the 1-particle reduced density matrix (1RDM) on cluster $J$ and $h_{pq}$, $\braket{pq}{rs}$, and $\expval{pq||rs}$ are the one-electron, simple two-electron, and antisymmetrized two-electron integrals, respectively.
The local cMF effective Hamiltonian (arising naturally from tracing out the remaining clusters) commutes with $\hat{N}$, $\hat{S}_z$, and $\hat{S}^2$, and as such the cluster states, $\ket{\alpha_I}$, automatically preserve  particle number and spin symmetries.  
Because the $\hat{H}^{eff}_I$ depends on all other clusters via the 1RDM, this must be solved self-consistently. 
The similarities between $\hat{H}^{eff}_I$ and the traditional Fock operator also extend to our ability to define a perturbation theory, as introduced in Ref. \citenum{Scuseria2015} and discussed later. 
For small clusters, this many-body transformation can simply be defined through the exact diagonalization (FCI) of $\hat{H}^{eff}_I$, although approximate eigenstates would be needed for larger clusters. 

In order to span the full Hilbert space of the global system, we must separately diagonalize Eq. \ref{eq:Heff} in all possible sectors of the cluster's local Fock space. 
The global states can then be represented  in the tensor product basis of cMF eigenstates:
\begin{equation}\label{eq:state}
    \ket{\Psi} = \sum_{\alpha}\sum_{\beta}...\sum_{\omega} c_{\alpha,\beta,\dots,\omega}\ket{\alpha_1}\ket{\beta_2}\dots \ket{\omega_N}
\end{equation}
where $c_{\alpha,\beta,\dots,\omega}$ is the coefficient tensor, and $\ket{\alpha_I}$ is an eigenvector of Eq. \ref{eq:Heff}.

The focus of this paper is to develop and test an excited state generalization of our  selected CI procedure (TPSCI)\cite{Abraham2020} which algorithmically builds a sparse approximation to Eq. \ref{eq:state}. 
The remaining theory section will be organized as follows: In Section \ref{states} we discuss clustering and how to generate initial cluster states by diagonalizing local Hamiltonians, Section \ref{matrix} provides details about the matrix element evaluation, and finally in Section \ref{alg} we explain in detail the steps of the TPSCI algorithm.

\subsection{Generating initial cluster states}\label{states}
\paragraph{cMF orbital optimization}
Because the cMF orbitals are optimized in addition to the cluster state coefficients, the final definition of the clusters are ultimately determined uniquely by the variational principle. However, a good initial guess is often necessary for the reliable orbital convergence  of cMF, similar to CASSCF.
There are several ways to generate an initial guess for  orbital clusters, and the best choice is often system dependent. 
The most straightforward (yet tedious) approach would be to localize the active space orbitals, and manually put them into clusters. 
Alternatively, one could use more automated strategies (based on graph theoretic algorithms or embedding inspired approaches).
Exploring and comparing these options will be the focus of future studies. 

In this work, we use a simple DIIS procedure to optimize the orbitals,\cite{Abraham2021,PULAY1980} where the orbital rotation gradient is taken as the error vector. 
For each set of orbitals, the  local CI coefficients are optimized self-consistently, and the optimized 1RDMs and 2RDMs are used to construct the new orbital gradient. 
As such, our optimization is a two-step procedure, consisting of an inner ``CI'' optimization, and an outer ``orbital'' optimization. 
Inclusion of the orbital Hessian and directly coupling orbital and CI coefficient degrees of freedom  would significantly improve the convergence.
However,  for this paper, the cMF is not the computational bottleneck, so we defer this to future work. 

\paragraph{Initial computation of the local cluster state basis}
The result of the cMF calculation is not only the variationally best tensor product state, but also a set of cluster-local effective Hamiltonians dressed in the  mean-field interactions of the other clusters.
We choose the eigenvectors of these effective Hamiltonians as our initial cluster state basis. 
While the default setting in our Julia implementation includes all possible electron numbers for a given cluster, functionality has been added to allow the user to define a net change ($\delta_e$) in particle number for each cluster.
This removes the cost associated with Fock sectors that will ultimately be insignificant in the final TPSCI wavefunction. 
Once the allowed particle number subspaces (Fock sectors) are defined, eigenvectors of the local cMF effective Hamiltonians are obtained in each cluster.
A total of $M$ (a user defined parameter)  eigenvectors are computed for each Fock sector and saved in memory as basis vectors.
Because these eigenvectors diagonalize a local Hamiltonian, all the local correlations are folded into the basis vectors.

\begin{figure}
    \centering
    \includegraphics[width=\linewidth]{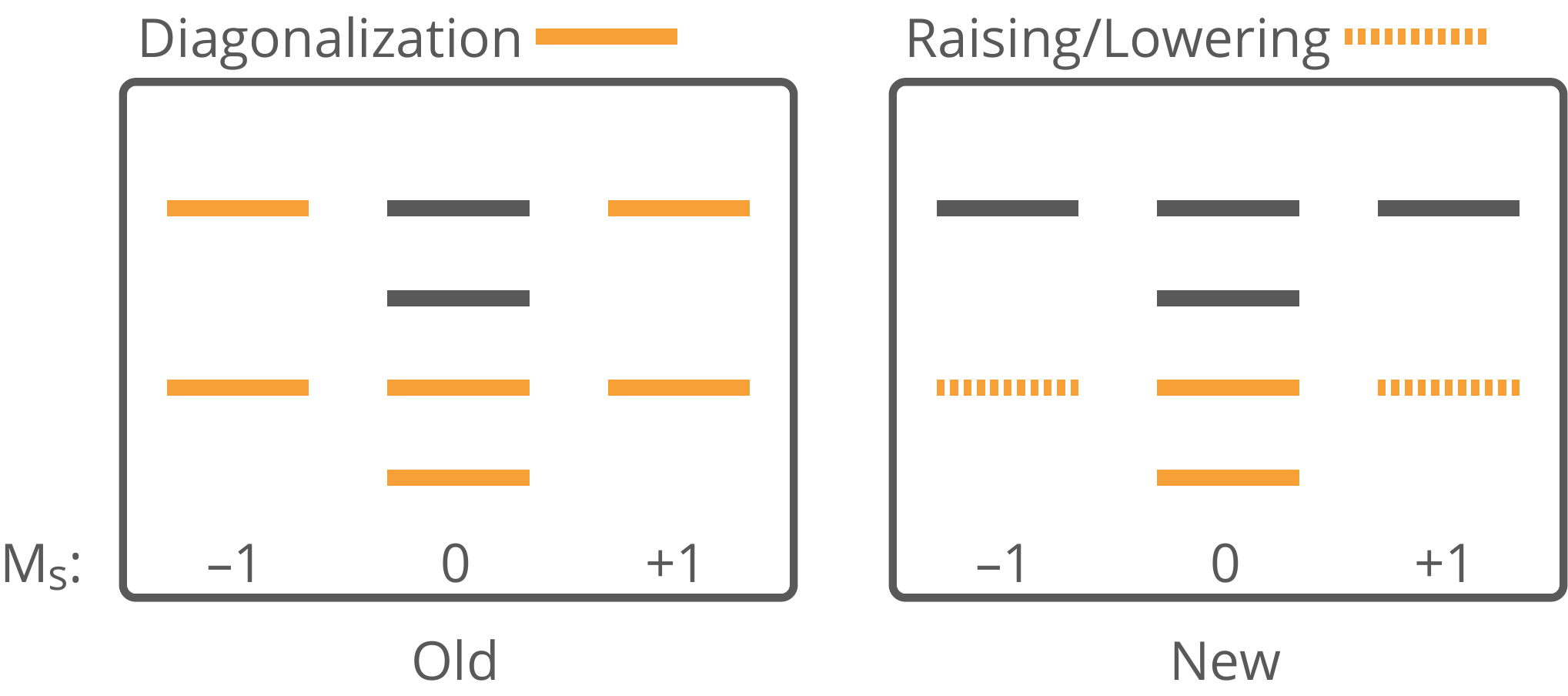}
    \caption{Comparison between the previous approach and the new approach for obtaining cluster states across different $m_s$ sectors. Grey (discarded states). Solid orange (states obtained via diagonalization). Dashed orange (states obtained by application of ladder operators). Here, we have an example of where a user selected to keep 2 states ($M=2$), and a cluster which has a singlet, triplet, singlet, triplet state ordering. In the old approach, our truncation would have incomplete treatment of the second triplet state, whereas the new approach is spin-complete.}
    \label{fig:ladder}
\end{figure}

\subparagraph{Spin Completeness of the basis}
In our initial TPSCI paper,\cite{Abraham2020} we computed the lowest $M$ states for each requested sector of Fock space,  treating each $m_s$ block independently. 
Because the different $m_s$ blocks have different dimensions, truncating with a fixed $M$ necessarily introduces spin-contamination into the global basis. 
While this was not a significant problem in our first paper focusing on ground states, for excited states spin-contamination 
can become more significant. 
To reduce this spin-contamination in the final TPSCI state (and to save both computational time and memory), in this newer implementation we simply generate the high- and low-$m_s$ components by directly applying the spin raising and lowering operators, $S^+$ and $S^-$, to the $m_s=0$ (even number of electrons) and $m_s=\tfrac{1}{2}$ (odd number of electrons) eigenstates. 
This ensures that all $m_s$ components are included for each cluster state computed such that $M$ state truncation does not break $\hat{S}_z$ symmetry. See Fig. \ref{fig:ladder} for a depiction of this process.
While this approach ensures that truncating $M$ does not create spin-contamination, there is still the possibility of creating spin-contamination in the SCI selection. 
This is a direct analogy to the situation with spin-contamination in determinant based selected CI codes.
We have added the ability to add the important TPS needed for achieving spin-complete wavefunctions by simply adding the dominant contributions from the $\hat{S}^2$ residual vector, $\ket{r_{\hat{S}^2}} = \left(\hat{S}^2 - \expval{\hat{S}^2}\right)\ket{\Psi^s}$.
However, for future work, we will likely attempt to generalize the recent work from Scemama and coworkers to also reduce the variational dimension.\cite{chilkuriChapterThreeSpinadapted2021}

\subsection{Matrix element evaluation}\label{matrix}
The diagonalization of $\hat{H}$ directly in the basis of tensor product states requires us to evaluate Hamiltonian matrix elements between arbitrary tensor product states. 
To save on time-complexity during the Hamiltonian evaluation, we precompute the representation of all the relevant local operators in the cluster basis. 
Following the relevant notation used in the ASD work,\cite{parkerCommunicationActiveSpace2014} we refer to them as $\Gamma$ tensors. 
To aid in the explanation of these tensors, we first introduce the standard electronic Hamiltonian in second quantization:
\begin{equation}
    \hat{H} = \sum_{pq} h_{pq}\hat{p}^\dagger \hat{q} + \frac{1}{2} \sum_{pqrs} \expval{pq\vert rs}  \hat{p}^\dagger \hat{q}^\dagger \hat{s} \hat{r}
\end{equation}
where $\hat{p}^\dagger$ and $\hat{q}$ are the fermionic creation and annihilation operators, and $h_{pq}$ and $\expval{pq\vert rs}$ are the one and two electron integrals, respectively. 
We partition the electronic Hamiltonian into one, two, three, and four cluster contributions, defined by the number of distinct clusters being acted upon:
\begin{align}\label{eq:ham}
    \hat{H} = & \sum_I \hat{H_I} + \sum_{I<J} \hat{H}_{IJ} \nonumber\\
    &+ \sum_{I<J<K} \hat{H}_{IJK} + \sum_{I<J<K<L}\hat{H}_{IJKL}
\end{align}
where $H_I$ has all creation and annihilation operators in cluster $I$, $H_{IJ}$ has operators in both clusters $I$ and $J$ and so on.
The full set of terms for each of these n-cluster interactions are included in the Supporting Information.
At most, we can only have four cluster interactions since we have at most four fermionic operators. 

Each of these terms will involve a contraction of the two electron integrals with the appropriate $\Gamma$ tensors. Therefore, we can precompute these terms and store them in a dictionary in memory for later access. 
For example, if we have a local state $\beta$ in cluster $I$ where operators $\hat{p}^\dagger$, $\hat{q}^\dagger$, and $\hat{r}$ act on cluster $I$, its associated $\Gamma$ tensor is the following:
\begin{equation}
{^I}\Gamma_{pqr}^{\beta'\beta} = \bra{\beta_I'}\hat{p}^\dagger \hat{q}^\dagger \hat{r} \ket{\beta_I}
\end{equation}
This is also an example of the largest rank $\Gamma$ tensor that our implementation will store in memory.  
For large clusters and large $M$ values (number of local cluster states), these can become the memory bottleneck. 
It is, in principle, possible to avoid the storage of these five-index tensors since they can only contribute to two-cluster terms, however we have not found the need yet.

These gamma tensors will be contracted with the integrals during the computation of each Hamiltonian matrix element.
For example, the following $\hat{H}_{IJ}$ term would provide the following contribution to the $\bra{\psi'} \hat{H}_{IJ} \ket{\psi}$ matrix element:
\begin{align}
\bra{\psi'} \hat{H}_{IJ} \ket{\psi} \leftarrow -(-1)^\chi \prod_{K\neq I,J}\delta_{\omega_K,\omega_K'}
\\\nonumber
\times\sum_{pqr\in I} \sum_{s \in J} \ip{pq}{rs} {}^I\Gamma_{pqr}^{\beta'\beta} {}^J\Gamma_{s}^{\gamma'\gamma}
\end{align}
where $\chi = \sum_{K=I}^{J-1} N_K$ and accounts for the sign by summing over the number of electrons in each cluster between the two active clusters, 
and $\delta_{\omega_K,\omega_K'}$ arises from the orthonormality between states $\omega$ and $\omega'$ on cluster $K$. 
There is an additional negative sign that arises from the anticommutator relationship when you switch the two annihilation operators $\hat{s}$ and $\hat{r}$ since the operators must be adjacent to the cluster they are acting upon. 

The orthonormality of the  cluster states creates sparsity in the Hamiltonian, such that we only need to compute contributions between tensor product states that have identical inactive clusters states.
Analogous to the Slater-Condon rules, only tensor product states that differ by less than 5 clusters can be coupled by the Hamiltonian. 

\begin{table}
    \centering
    Table of TPSCI parameter definitions
    \begin{tabular}{c|p{3in}}
    \hline\hline
         $N$ & Number of clusters \\\hline
         {\tt R} & Number of global eigenvectors requested \\\hline
         $M$ &  Maximum number of cluster states in any given sector of Fock space for any cluster\\\hline
         $\delta_e$  & Range of Fock sectors for each cluster to include. For example, if cluster $I$ has 10 electrons in the cMF reference, then compute cluster states for $10-\delta_e$ $\rightarrow$ $10+\delta_e$\\\hline
         $\epsilon_\text{CIPSI}$ &   Threshold for discarding first-order TPS coefficients. Coefficients larger than this value will be included in the variational space.\\\hline
         $\epsilon_\text{FOIS}$ & Threshold for screening when computing the first-order interaction space. Values larger than this will be included when computing the first-order wavefunction.\\
         \hline\hline
    \end{tabular}
    \caption{Table of definitions of parameters used to define a TPSCI calculation}
    \label{tbl:definitions}
\end{table}
\begin{figure}
    \includegraphics[width=\linewidth]{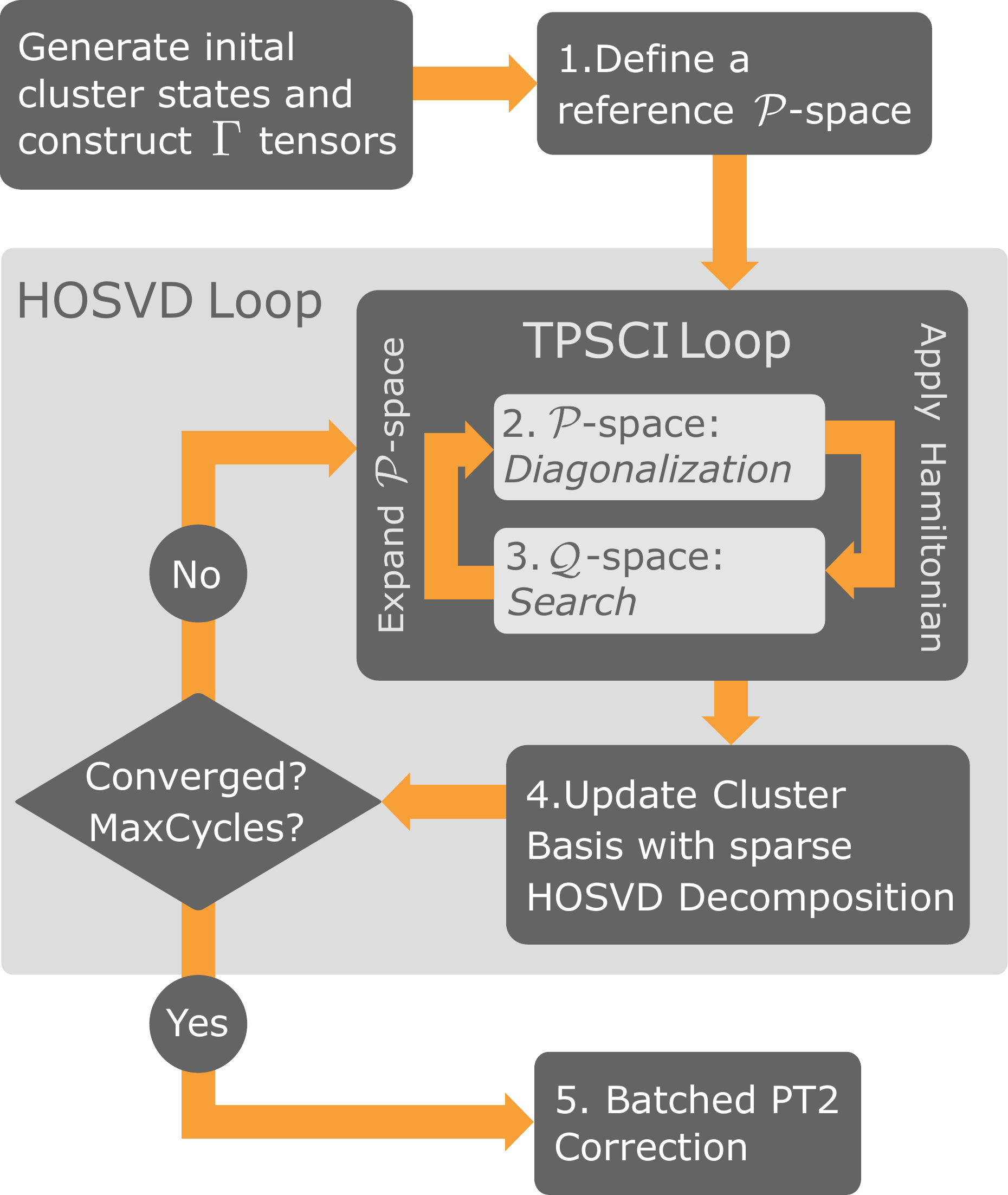}
    \caption{Flow chart of the TPSCI algorithm including the HOSVD loop.}
    \label{fig:tpsci}
\end{figure}
\subsection{Algorithm}\label{alg}
We start by first listing the overall steps for the TPSCI algorithm (which can also be seen in Figure \ref{fig:tpsci}), then follow with a more detailed discussion of each step.
We also include a table of the required user-defined parameters for a TPSCI calculation Table \ref{tbl:definitions}.

Steps of a TPSCI calculation for computing {\tt R} states:
\begin{enumerate}
    \item Define a reference $\mathcal{P}$-space with dimension of at least {\tt R}. (Sec. \ref{ref_p})
    \item Diagonalize the Hamiltonian in the $\mathcal{P}$-space and collect lowest {\tt R} eigenstates. (Sec. \ref{diag})
    \item Search $\mathcal{Q}$-space perturbatively and expand $\mathcal{P}$-space. If converged, continue, else return to step 2. (Sec. \ref{fois})
    \item Update cluster basis with sparse higher-order singular value decomposition (HOSVD) decomposition. If converged, continue, else return to step 2. (Sec. \ref{tucker})
    \item Compute a state specific  PT2 energy correction. (Sec. \ref{pt2})
\end{enumerate}

\subsubsection{Define a Reference $\mathcal{P}$ Space}\label{ref_p}
For ground state TPSCI calculations, the cMF wavefunction often serves as a sufficient initial $\mathcal{P}$ space.
However, for excited states it is often helpful to specify an initial $\mathcal{P}$ space that qualitatively describes the target states.  

If the system were to be fully decoupled such that there were no interactions between clusters, 
then the full Hamiltonian would be diagonal in the TPS basis. 
Additionally, the low-energy spectrum would be dominated by ``excitonic'' states, those states where every cluster is in its ground state except for a single (or pair) which is excited. 
However, as the clusters become more strongly interacting, the low energy spectrum can develop greater weight on higher exciton-rank tensor products. 
For weakly to moderately interacting clusters, the excitonic basis provides a qualitatively correct description of the target excited states, and thus is an excellent initial $\mathcal{P}$ space for starting the TPSCI procedure. 
The single excitonic basis for a given cluster can be written as:
\begin{equation}
\ket{\psi_{\lambda_{L}}} =   \ket{0_{I},0_{J},..\lambda_{L}, ..0_{N}}
\end{equation}
where cluster $L$ is in its singly excited state $\lambda$.
For very weakly interacting systems, 
one would expect the low-energy states to be primarily represented as linear combinations of these single excitonic states. 
For comparative purposes, we will refer to such a method as TPS-single exciton (TPS-SE).
This is equivalent to the so-called Block correlated CI method described by Li and coworkers.\cite{liBlockcorrelatedCoupledCluster2004} 
Although the TPS-SE results will not generally be accurate since it lacks all  interactions with higher excited configurations (e.g. the charge transfer excitations), 
the TPS-SE method provides a very effective way to initialize the TPSCI calculation with a qualitatively correct initial $\mathcal{P}$ space. 
Further, for situations where we expect biexcitons to contribute to the final wavefunctions (see Sec. \ref{sec:sf}), the user can also directly add these configurations to the starting wavefunction. 

We provide a comparison of the TPS-SE with TPSCI for one of the systems we studied (P1) in the Supporting Information which demonstrates that  one does generally need to go beyond TPS-SE for accurate excited states.

\subsubsection{$\mathcal{P}$-space: Diagonalization}\label{diag}
Once the variational space is defined, we build the Hamiltonian from Eq. \ref{eq:ham} in the $\mathcal{P}$ space and diagonalize. 
As described above, the required matrix element evaluation is much more expensive than traditional Slater determinant methods due to two main reasons:
i) the loss of sparsity of the Hamiltonian matrix, and 
ii) the need to contract the integrals with the precomputed $\Gamma$ tensors mentioned in \ref{matrix}.
Because the Hamiltonian matrix storage usually constitutes a memory bottleneck,  we have implemented the option for either a full matrix build or a matrix-vector product build for use in a Krylov solver. 
However, while the matrix-vector algorithm significantly reduces the memory requirements, it is much slower because it recomputes the matrix elements for each Lanczos iteration. 
As such, if allowed by memory, our current implementation defaults to the full Hamiltonian matrix build. 
After we build and diagonalize the Hamiltonian, we have a set of variational states that are a sum of tensor product states, $\ket{\mathcal{P}^s} = \sum_i c_i^s\ket{\mathcal{P}_i}$ and a variational energy, $E_0$.

\subsubsection{$\mathcal{Q}$-space: Search}\label{fois}
To obtain the first-order interacting space (FOIS), we calculate the action of the Hamiltonian on the set of tensor product states, $\{\mathcal{P}_i\}$, in the current variational space. 
\begin{align}
    \ket{\sigma_j^s} =& \sum_{i}\dyad{\mathcal{Q}_j}\hat{H}
    \ket{\mathcal{P}_i}c_i^s\nonumber\\
    =& \ket{\mathcal{Q}_j}b_j^s
    \end{align}
The states, $\ket{\mathcal{Q}_j}$, run over all tensor product states that can be reached by the Hamiltonian from the current variational space, excluding the variational space.
Since the Hamiltonian is not sparse in the TPS basis, the action of the Hamiltonian on the TPS states can become very costly.  Therefore, we have implemented a series of screening and prescreening techniques based on a user-defined threshold, $\epsilon_\text{FOIS}$, 
where we delete components $\ket{\mathcal{Q}_j}$ if $\displaystyle\max_s\vert b_j^s\vert\leq \epsilon_\text{FOIS}.$
We then collect the resulting non-negligible configurations that lie in the $\mathcal{Q}$ space. 

Consistent with the original Slater determinant Configuration Interaction Perturbatively Selected Iteratively (CIPSI)\cite{Huron1973} method, we compute the first order correction to our current variational state(s) to determine which new degrees of freedom should be added to our variational space. 
In our work, we use a generalization of the Barycentric Moller Plesset\cite{Huron1973} (MP) perturbation theory using the cMF effective Hamiltonian (Eq. \ref{eq:Heff}), which is explicitly described in Appendix \ref{app:pt}.

Once the first order coefficients are computed for each state, $\ket{\mathcal{P}^s}$,  
\begin{align}
c_j^{s(1)} = \sum_{i}\frac{
\mel{\mathcal{Q}_j}{\hat{H}}{\mathcal{P}_i}
c_i^{s(0)}
}{\Delta E^{(0)}_j}    
\end{align}
any $\mathcal{Q}$ space configuration with a perturbative coefficient greater than $\epsilon_\text{CIPSI}$ ( $\displaystyle\max_s|c_j^{s(1)}| > \epsilon_\text{CIPSI}$) is added to the $\mathcal{P}$ space.
\footnote{Note that the threshold used in the earlier TPSCI work\cite{Abraham2020} was pruning the probabilities instead of the coefficients like we do in this work.}
If no additional TPS states are added to the variational space, the TPSCI protocol is considered converged.

\subsubsection{Update Cluster Basis with HOSVD Decomposition}\label{tucker}
Once the TPSCI wavefunction has converged (i.e. no additional TPS states are required in the variational space), we can optionally update the cluster basis using a quantum number-preserving Tucker decomposition called a higher-order SVD decomposition (HOSVD) 
\begin{equation}\label{eq:hosvd}
    \mathcal{T}_{\alpha,\beta,\dots,\gamma} = \mathcal{C}_{i,j,\dots,d}U_{\alpha,i}U_{\beta,j} \cdots U_{\gamma,d}
\end{equation}
where $\alpha,\beta,\dots,\gamma$ are each specific to a cluster,
and $\mathcal{C}_{i,j,\dots,d}$ is the core tensor which is formed by a change of basis from $\alpha$ to $i$, $\beta$ to $j$ etc.
Because we are using the HOSVD to only rotate the cluster basis, and \textit{not} truncate the space,\cite{Abraham2020, mayhall_hosvd} each $U$ is a unitary matrix in the vector space of its specified  cluster. 
These unitary matrices are local many-body rotations which can be directly obtained from individual singular value decompositions (SVD) along the associated axis, e.g.
\begin{equation}
    \mathcal{T}_{\alpha,\beta\dots\gamma} = U_{\alpha,i}\Sigma_i V_{i, \beta\dots\gamma},
\end{equation}
or equivalently, by diagonalizing the cluster reduced density matrix (cluster-RDM) which is obtained by tracing out the remaining clusters from the converged TPSCI wavefunction. 
\begin{equation}
    \rho_{\alpha \alpha'} = c(\alpha, \beta, \dots, \gamma)c(\alpha', \beta, \dots, \gamma)
\end{equation}
where $c(\alpha, \beta, \dots, \gamma)$ is the TPS coefficient vector. 
We note that, in practice, we want to preserve certain local quantum numbers (particle number and spin projection). 
As such, we only block-diagonalize the cluster-RDM within each quantum number subspace. 
This ensures that the global wavefunction retains proper eigenstates of both $\hat{N}$ and $\hat{S}_z$.


When moving to a multi-state problem, there are various ways to complete this HOSVD to obtain the tucker factors ($U$).
One option is to decompose each state into its own basis. 
However this state-specific approach would be extremely complex, making it difficult to reliably compute energy differences and transition properties between states. 
Instead, we compute a single global basis in a state-averaged way.
To create this global basis we simply average the cluster-RDMs from each TPSCI eigenvector:
\begin{equation}
    \rho_{\alpha \alpha'} = \tfrac{1}{\tt R} \sum_s\sum_{\beta, \dots, \gamma}c(\alpha, \beta, \dots, \gamma)^s c(\alpha', \beta, \dots, \gamma)^s
\end{equation}
where $s$ is denoting the state.
We can then diagonalize $\rho_{\alpha \alpha'}$ to obtain the tucker factors for cluster $\alpha$:
\begin{equation}
    \rho_{\alpha \alpha'} = U_{\alpha,i} \space g_i U_{\alpha', i}.
\end{equation}
We view the use of the HOSVD as optional, analogous to the use of natural orbitals in conventional Slater determinant selected CI calculations. As such, it is obtained iteratively, where cheaper calculations provide states which are decomposed to produce more compact representations for subsequent calculations with tighter thresholds, $\epsilon_\text{CIPSI}$.
We refer to this computational protocol of systematically  tightening the thresholds after one (or more) HOSVD steps as ``HOSVD bootstrapping'' in the results section. 


\subsubsection{Batched PT2 Energy Correction}\label{pt2}
Even though the Selected-CI algorithm captures most of the static correlation with the CI expansion, it does not capture the dynamic correlation efficiently enough to produce near FCI accuracy results.
The inclusion of the missing dynamic correlation is usually carried out using a state-specific PT2 correction.
It is important to note that we are only referring to dynamical correlation inside of the active space, out-of-active space correlations would still need to be accounted for (either through downfolding, PT2, or adiabatic connection type schemes) to enable direct comparison to experiment.
 As mentioned in Sec. \ref{fois}, in the cMF basis, we choose a Barycentric Moller Plesset\cite{Huron1973} (MP) type partitioning in this work.
Whereas computing the first-order wavefunction can quickly become a memory bottleneck due to the vast size of the $\mathcal{Q}$ space, the energy computation has no inherent memory demand. 

For computing the PT2 energy correction, 
we have implemented a parallelized batched algorithm, where we compute a small segment, or batch, of the first order wavefunction, then contract it to evaluate the energy, discarding the state before moving to the next segment. 
Our current implementation batches over what we refer to as {\tt FockConfig}'s, or unique distributions of particles across clusters.
This approach is analogous to the determinant based approach described in Ref. \citenum{tubmanEfficientDeterministicPerturbation2018}.
While this does offer system-dependent speedups, the scaling is far from optimal. 
The reason is that by parallelising over Fock space configurations, 
we have a  rather poor load balancing due to the fact that some Fock space configurations have many more configurations than others. Improvements to our batching will be the focus of future work. 


\section{Results and Discussions}
We investigate the efficiency of the TPSCI approach for excited states by mainly focusing on polycyclic aromatic hydrocarbons (PAH).
These systems have been chosen for three reasons:
i) they provide a straightforward approach to orbital clustering, allowing us to defer the more complicated clustering patterns to focused future work, 
ii) we have already begun to understand the ground state behavior in our previous paper\cite{Abraham2020}, and 
iii) because they are chemically interesting  in terms of novel material in synthesizing chiral nanographenes,\cite{Reger2021} twisted carbon nanobelts,\cite{Wei2021} and carbon-based electronic devices \cite{Fan2021} etc.. 
Benchmarking on a wider variety of chemical systems will be the focus of follow up papers.

The first few systems (Section \ref{sec:ex_pah}) constitute  a set of $\pi$ conjugated systems which can be grouped into clusters of six orbitals which simply differ in their connectivity.
The last example is a tetracene tetramer, 
which is non-covalently bound and supports interesting multiexcitonic states.
For all systems we compute accurate estimates of both the ground state and a large number of excited states. 
\begin{figure}
    \includegraphics[width=\linewidth]{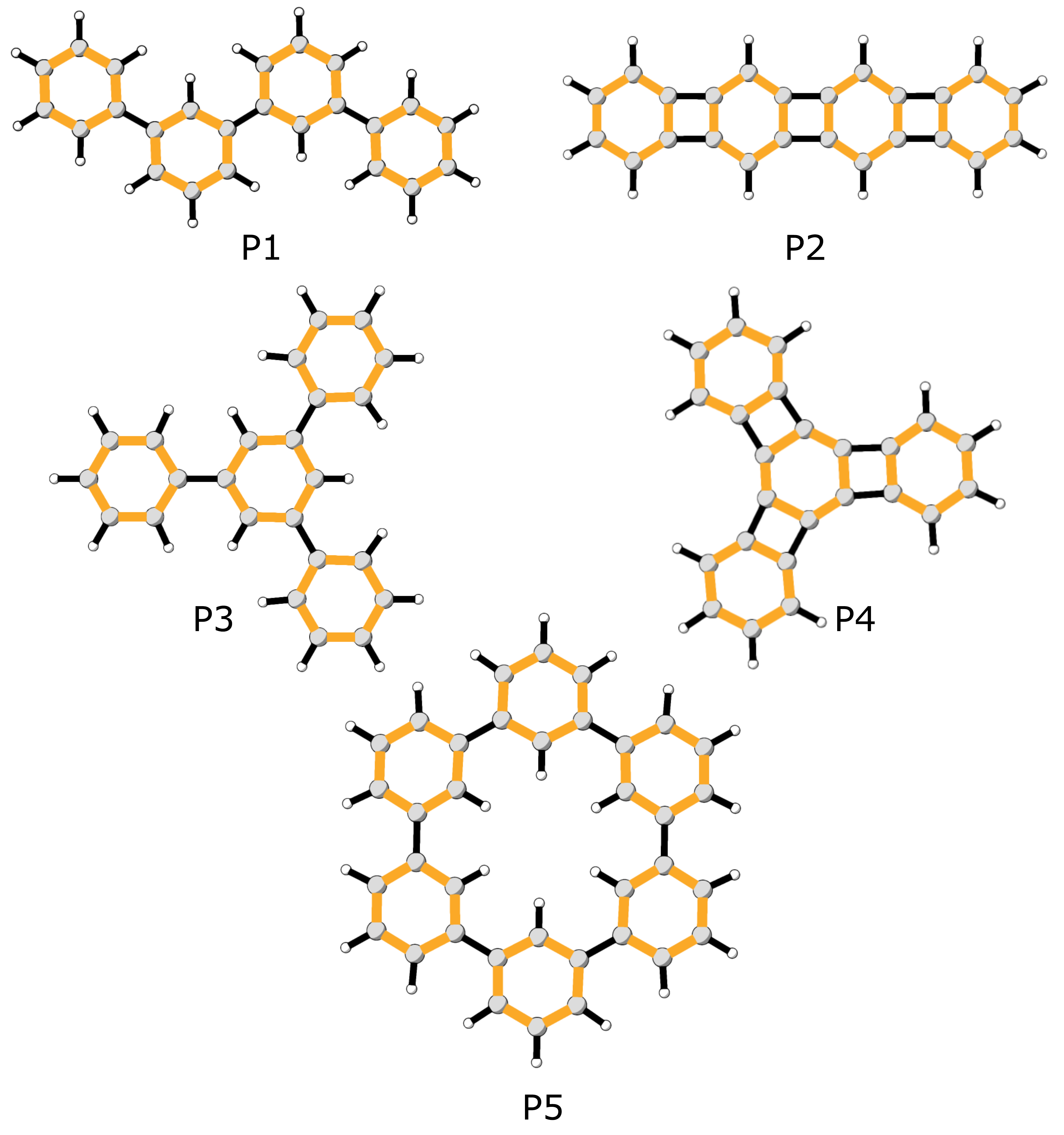}
    \caption{PAH systems used for the excitation energies. Each gold highlighted region corresponds to a separate cluster.}
    \label{fig:ex_system}
\end{figure}

For the PAH systems, we use geometries optimized at the B3LYP/cc-pVDZ\cite{dunning1989a} level of theory.
The active space for each PAH system P1-P5 is generated by extracting the molecular orbitals that have significant $2p_{z}$ atomic orbital character then localized using the Boys\cite{boys} localization method.
We use the 6-31G*\cite{ditchfield1971a} basis for the singlet fission tetracene tetramer calculations.
The active space for the tetracene tetramer system is generated by first obtaining a set of natural orbitals obtained by diagonalizing a state-averaged CIS density (CIS-NO),\cite{shuConfigurationInteractionSingles2015} then these are localized using the Pipek-Mezey\cite{PM} localization method.
All semi-stochastic heat-bath CI (SHCI) calculations were performed with Arrow\cite{arrow1,arrow2,arrow3,arrow4}.
The integrals for all the calculations were generated using the PySCF package,\cite{pyscf2020}
and the cMF and TPSCI calculations were performed with our open source Julia\cite{bezanson2017julia} packages {\tt ClusterMeanField.jl}\cite{clustermeanfield} and  {\tt FermiCG}.\cite{fermicg}
The geometries for all the systems are included in the Supporting Information.

It is important to note that the calculations reported in this section are not capable of being directly compared to experiment. While we believe they are highly accurate \textit{inside} the active spaces, more work is needed to provide direct comparisons with experiment. In particular, it will be necessary to include the missing sigma-bond correlation,\cite{angeliNatureIonicExcited2009} dynamic correlation outside of the active space, larger basis set effects, and vibronic effects to make sure our calculations are directly comparable to experiment.

We note that the thresholding used in the original ground state TPSCI work\cite{Abraham2020} pruned by using the probability and hence was square of the $\epsilon_\text{CIPSI}$ in this work.
The current work prunes on the absolute value of the first order coefficients to be more consistent with other selected CI codes. 

\begin{figure*}
    \includegraphics[width=0.9\linewidth]{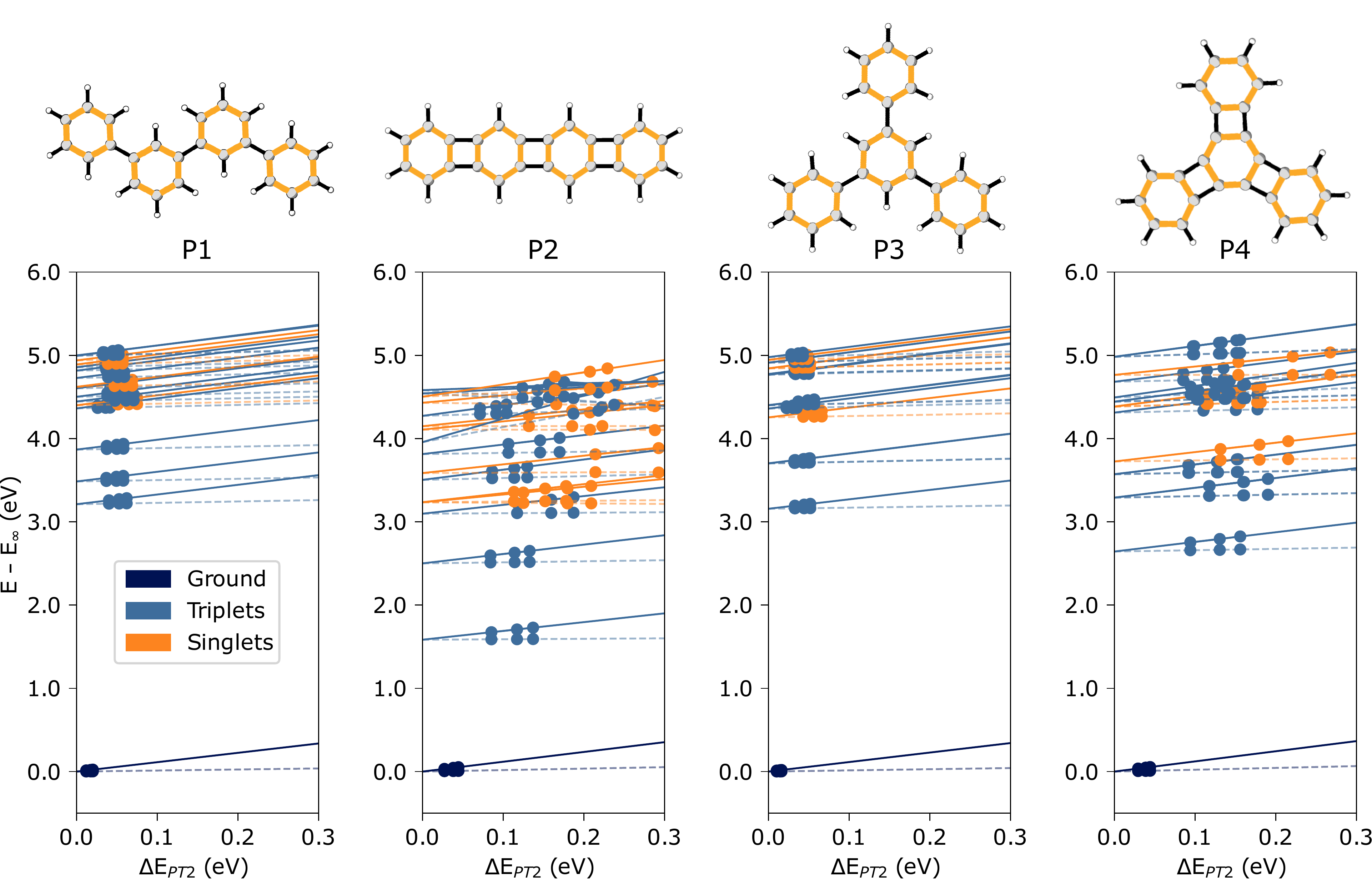}
    \caption{Extrapolation of the ground state and 16 excited states for the medium sized PAH systems studied using TPSCI (with HOSVD bootstrapping). After the bootstrapping, the TPSCI wavefunction coefficients are clipped at the larger thresholds to obtain the additional points in the extrapolation to plot against the PT2 enery correction with root tracking. All energies are shifted by the extrapolated ground state TPSCI energy so the ground state converges to 0 eV.
    Variational energy fit (solid lines).
    PT2 energy fit (dashed lines).
    ($\epsilon_\text{CIPSI} = n \times 10^{-4}$ with n=4,6,8 for P1, P3, and P4. $\epsilon_\text{CIPSI} = n \times 10^{-4}$ with n=6,8,10 for P2)}
    \label{fig:ex_data}
\end{figure*}

\subsection{PAH systems}{\label{sec:ex_pah}}
We present four medium sized PAH systems (P1-P4) and one larger system (P5) (Figure \ref{fig:ex_system}).
Taking the $\pi$ space as the active space, P1-P4 has an active space of size 24 electrons in 24 orbitals (4 clusters) and P5 has an active space of 36 electrons in 36 orbitals (6 clusters).
Considering that the low-energy excited states of benzene consist of one singlet state and three triplet states, in our calculations on P1-P4, we compute 16 total excited states, while for P5 we compute 24 excited states (i.e., four states per cluster).

\subsubsection{Smaller PAH systems (P1-P4)}
In Figure \ref{fig:ex_data}, we present the extrapolation of the ground and 16 excited states for systems P1-P4 using TPSCI where the ground state is colored navy blue, triplets are colored blue, and singlets are colored orange.

As is commonly done in selected CI calculations, we assume a linear relationship between  the PT2 energy correction and variational energy (i.e. the larger the cutoff, the cheaper the selected CI calculation, therefore more energy correction will be required). 
The extrapolated results in Figure \ref{fig:ex_data} were computed by first converging to the tightest $\epsilon_\text{CIPSI}$ possible ($4\times 10^{-4}$ for P1, P3, and P4 and $6\times 10^{-4}$ for P2) through the HOSVD bootstrapping approach. 
The additional cheaper points for extrapolation were obtained by deleting TPS's with a coefficient smaller than a specified epsilon value, and then recomputing the eigenvectors and PT2 corrections in these successively smaller variational spaces.
We note that the same cluster basis is used for each point in the extrapolations, i.e., we don't perform any additional HOSVD for the extrapolation points. 
This allows us to track states and monitor if any root flips across extrapolation points.
The point where the extrapolated lines cross the y-axis  (i.e., where the variational energy is predicted to have a zero PT2 correction) is our best estimate of the FCI energies. 
Therefore, the closer the variational energies are to the extrapolated result (y-axis), the more reliable the calculation.

For each of the  P1-P4 systems, the ground state variational estimate converges much faster than the excited states as seen from Figure \ref{fig:ex_data}.
In future work, we plan to investigate the use of iso-PT2 selection schemes to help improve the convergence of excited states.\cite{dashExcitedStatesSelected2019, dashTailoringCIPSIExpansions2021}

The TPSCI results for the singly connected systems, P1 and P3, converge much faster than the P2 and P4 systems.
This is to be expected given the fact that each cluster is connected by two bonds instead of one,
leading states in the P2 and P4 systems to develop significantly more inter-cluster entanglement.

Overall, we see that qualitatively, the low energy electronic structure of the clusters is retained when the system is more weakly coupled, than otherwise. For instance, for P1, P3, and P4 there are three triplets for every singly excited benzene unit (4 singlets, and 12 triplets). 
This is the same ratio that is found in the isolated benzene structure. 
In contrast, for the P2 system, we observe 7 singlets and 9 triplets within the lowest 16 states. 
We interpret this increase in singlet contribution to arise from the increased interactions between the clusters, which provides more ability for the electronic structure to delocalize between clusters.

Although P4 also has clusters which are connected by two bonds, the non-linear geometry prevents the qualitative reorganization of the electronic structure such that there are still 4 singlets and 12 triplets. 
Further, unlike P2, the singlet-triplet gap is not significantly lowered compared to P1 or P3. 

We note that in the P2 extrapolated graph, we observe a very steep slope for one of the states around 4.0 eV.
This could indicate that this state was not converged tightly enough for extrapolation.
Alternatively, it might have arisen from the manner in which we apply perturbation theory. 
As mentioned above, we are currently using a non-degenerate PT2 formalism, which can create problems in cases of near degeneracy. 
In follow up work, we plan on implementing a quasi-degenerate formalism,\cite{lowdinNoteQuantumMechanical1951, 
lowdinStudiesPerturbationTheory1962, 
gershgornApplicationPerturbationTheory1968, 
angeliQuasidegenerateFormulationSecond2004, shavittQuasidegeneratePerturbationTheories2008, sharmaQuasidegeneratePerturbationTheory2016a} 
following a strategy similar to our recent work,\cite{mayhallIncreasingSpinflipsDecreasing2014, mayhallQuasidegenerate2ndorderPerturbation2014} to better understand the current results, and to safeguard against such issues in the future. 

\subsubsection{Larger PAH (P5)}
\begin{figure}
    \includegraphics[width=0.99\linewidth]{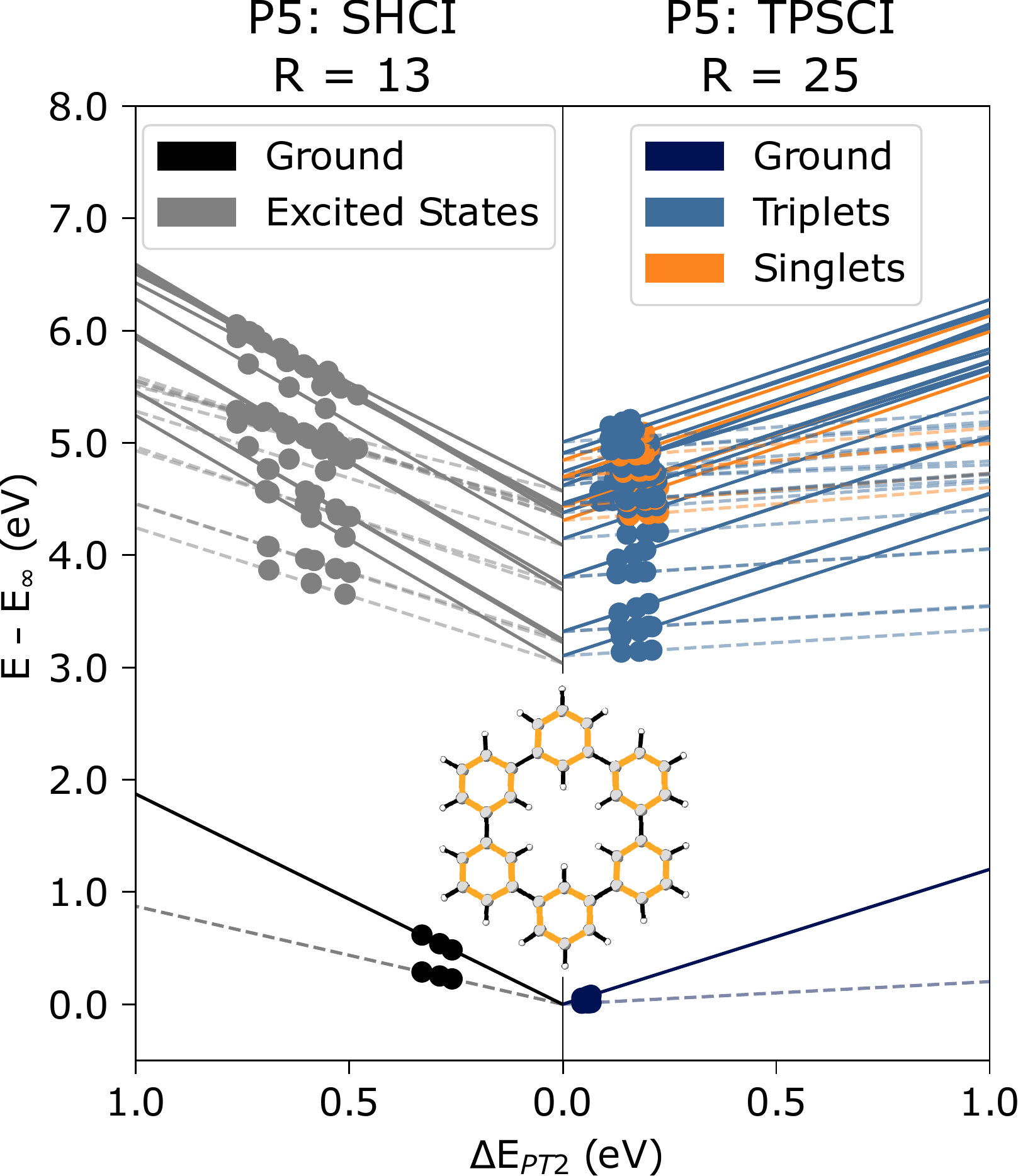}
    \caption{Extrapolation for the P5 molecule using the SHCI and TPSCI methods respectively. R denotes the number of roots: 13 for SHCI,  and 25 for TPSCI. (TPSCI $\epsilon_\text{CIPSI} = n \times 10^{-4}$ with n=4,6,8 and SHCI $\epsilon_\text{CIPSI} = n \times 10^{-5}$ with n=5,7,10)}
    \label{fig:p5}
\end{figure}

As a larger example of a $\pi$-conjugated system, we also consider P5, which has an active space of 36 electrons in 36 orbitals that is partitioned into 6 clusters.
Similar to before, we expect three triplets and one singlet for every singly excited cluster, giving a total of 25 states. 
We present the extrapolations of both TPSCI and semi-stochastic heat bath CI (SHCI) in Figure \ref{fig:p5}.
The linear extrapolation 
has been shown previously in literature to generate overestimated energies.\cite{Holmes2017}
A quadratic fit is recommended in these cases, but for comparison we use a linear fit for both methods.

In order to label the eigenstates, we compute the expectation value of $\hat{S}^2$ for each of the TPSCI states. 
Although the TPSCI results are rather tightly converged, 
nearly-degenerate states can mix arbitrarily,  
leading to a few instances of  non-trivial spin-contamination. 
However, the extent of this is generally small enough such that it doesn't prevent us from labeling the states.%
\footnote{We could also approximate the eigenstates by diagonalizing $\hat{H} + \alpha\hat{S}^2$ for some small value of $\alpha$ to break the near degeneracies between states of different spin, however we have not done this here.
}
Using SHCI, we were not able to compute all 25 states because the variational space grew too large to fit in memory. 
The largest calculation we were able to obtain was for 13 roots. 
Further, we didn't have access to $\expval{\hat{S}^2}$ values for SHCI, so we were not able to label the resulting states, and thus they are simply left grey in Fig. \ref{fig:p5}.


In addition to the plots in Fig. \ref{fig:p5}, we also present these results in Table \ref{tbl:p5}.
Here,  we report the variational excitation energies ($\omega_\text{Var}$),  magnitude of the PT2 correction to the excitation energies ($\Delta\omega_\text{PT2} = \omega_\text{PT2} - \omega_\text{Var}$), and extrapolated excitation energies ($\omega_\infty$).
To better highlight the accuracy of the perturbatively corrected results, we also present the extrapolation corrections ($\Delta \omega_\infty = \omega_\infty-\omega_\text{PT2}$) to the excitation energies.

For all excited states computed, the TPSCI variational energy is closer to its extrapolated result than the corresponding variational HCI result.
This is a consequence of folding in local correlations directly into the TPS basis. 
Not only do the TPSCI results have smaller PT2 corrections ($\Delta\omega_\text{PT2}$) compared to SHCI, but more importantly, the extrapolation correction is significantly smaller than the PT2 correction for each state, 
$\Delta\omega_\text{PT2} > \Delta\omega_\infty$.
In contrast, this isn't the case for the SHCI results, where the extrapolation corrections are consistantly larger than the PT2 corrections. 
For all excitation energies, the magnitude of $\Delta\omega_\text{PT2}$ for  SHCI  is around a factor of three times that of TPSCI.

The fact that the TPSCI variational (and perturbative) results are closer to the extrapolated values lends greater confidence to the extrapolated values. 
This is extra important in situations where different methods yield extrapolations that differ non-trivially, as seen in Fig. \ref{fig:p5}. 
While the overall features are similar between SHCI and TPSCI, the extrapolated values differ by a non-negligible amount (up to around 100 meV). 
Because of the fact that our extrapolation is smaller, we expect that the TPSCI extrapolations are closer to the exact FCI results than are the SHCI extrapolations.\footnote{
We have tested increasing the number of cluster states and found that our TPSCI results didn't change significantly, ruling out that any truncation effects were at play.
}

\begin{table}[]
\begin{center}
\caption{Excitation energies (eV) and wavefunction dimension for the most accurate calculation reported for the P5 system using TPSCI and SHCI ($\epsilon_\text{CIPSI}=4\times 10^{-4}$ for TPSCI and $\epsilon_\text{CIPSI}=5\times 10^{-5}$ for SHCI). Lineally extrapolated results obtained from PT2 energy corrections. $\omega_\text{Var}$ is the variational excitation energy, $\Delta\omega_\text{PT2}$ is the PT2 energy correction to the excitation energy, $\Delta \omega_\infty$ is the extrapolation correction, and $\omega_\infty$ is the extrapolated excitation energy.}
\label{tbl:p5}
\begin{tabular}{|c|c c c |c| c c c |c|}
\hline 

 & \multicolumn{4}{c|}{TPSCI} & \multicolumn{4}{c|}{SHCI} \\ 
 & \multicolumn{4}{c|}{Dimension: $112,788$} & \multicolumn{4}{c|}{Dimension: $1,741,084$} \\ \hline
State & $\omega_\text{Var}$  & $\Delta\omega_\text{PT2}$    & $\Delta \omega_\infty$ &  $\omega_\infty$ & $\omega_\text{Var}$  & $\Delta\omega_\text{PT2}$    & $\Delta \omega_\infty$ &  $\omega_\infty$ \\ \hline
1	&	3.22	&	-0.09	&	-0.02	&	3.10	&	3.68	&	-0.25	&	-0.39	&	3.04	\\
2	&	3.43	&	-0.09	&	-0.02	&	3.32	&	3.86	&	-0.24	&	-0.37	&	3.25	\\
3	&	3.43	&	-0.09	&	-0.02	&	3.32	&	3.93	&	-0.27	&	-0.43	&	3.22	\\
4	&	3.91	&	-0.08	&	-0.02	&	3.80	&	4.37	&	-0.25	&	-0.38	&	3.74	\\
5	&	3.91	&	-0.08	&	-0.02	&	3.80	&	4.40	&	-0.27	&	-0.44	&	3.69	\\
6	&	4.28	&	-0.11	&	-0.03	&	4.15	&	4.82	&	-0.30	&	-0.44	&	4.09	\\
7	&	4.46	&	-0.11	&	-0.04	&	4.31	&	4.94	&	-0.22	&	-0.34	&	4.38	\\
8	&	4.52	&	-0.04	&	-0.01	&	4.46	&	5.00	&	-0.26	&	-0.40	&	4.34	\\
9	&	4.52	&	-0.11	&	-0.03	&	4.38	&	5.03	&	-0.31	&	-0.29	&	4.43	\\
10	&	4.58	&	-0.11	&	-0.03	&	4.44	&	5.07	&	-0.27	&	-0.40	&	4.39	\\
11	&	4.58	&	-0.11	&	-0.04	&	4.44	&	5.11	&	-0.30	&	-0.47	&	4.34	\\
12	&	4.59	&	-0.10	&	-0.03	&	4.45	&	5.15	&	-0.29	&	-0.29	&	4.57	\\
13	&	4.59	&	-0.10	&	-0.03	&	4.46	&	--	&	--	&	--	&	--	\\
14	&	4.74	&	-0.08	&	-0.04	&	4.62	&	--	&	--	&	--	&	--	\\
15	&	4.74	&	-0.08	&	-0.04	&	4.62	&	--	&	--	&	--	&	--	\\
16	&	4.77	&	-0.09	&	-0.01	&	4.67	&	--	&	--	&	--	&	--	\\
17	&	4.78	&	-0.09	&	-0.01	&	4.68	&	--	&	--	&	--	&	--	\\
18	&	4.83	&	-0.10	&	-0.03	&	4.70	&	--	&	--	&	--	&	--	\\
19	&	4.83	&	-0.10	&	-0.03	&	4.70	&	--	&	--	&	--	&	--	\\
20	&	4.87	&	-0.09	&	-0.03	&	4.74	&	--	&	--	&	--	&	--	\\
21	&	4.97	&	-0.09	&	-0.03	&	4.85	&	--	&	--	&	--	&	--	\\
22	&	5.00	&	-0.07	&	-0.02	&	4.91	&	--	&	--	&	--	&	--	\\
23	&	5.00	&	-0.07	&	-0.02	&	4.91	&	--	&	--	&	--	&	--	\\
24	&	5.09	&	-0.07	&	-0.02	&	5.01	&	--	&	--	&	--	&	--	\\ \hline
\end{tabular}
\end{center}
\end{table}
\begin{figure*}
    \includegraphics[width=0.85\linewidth]{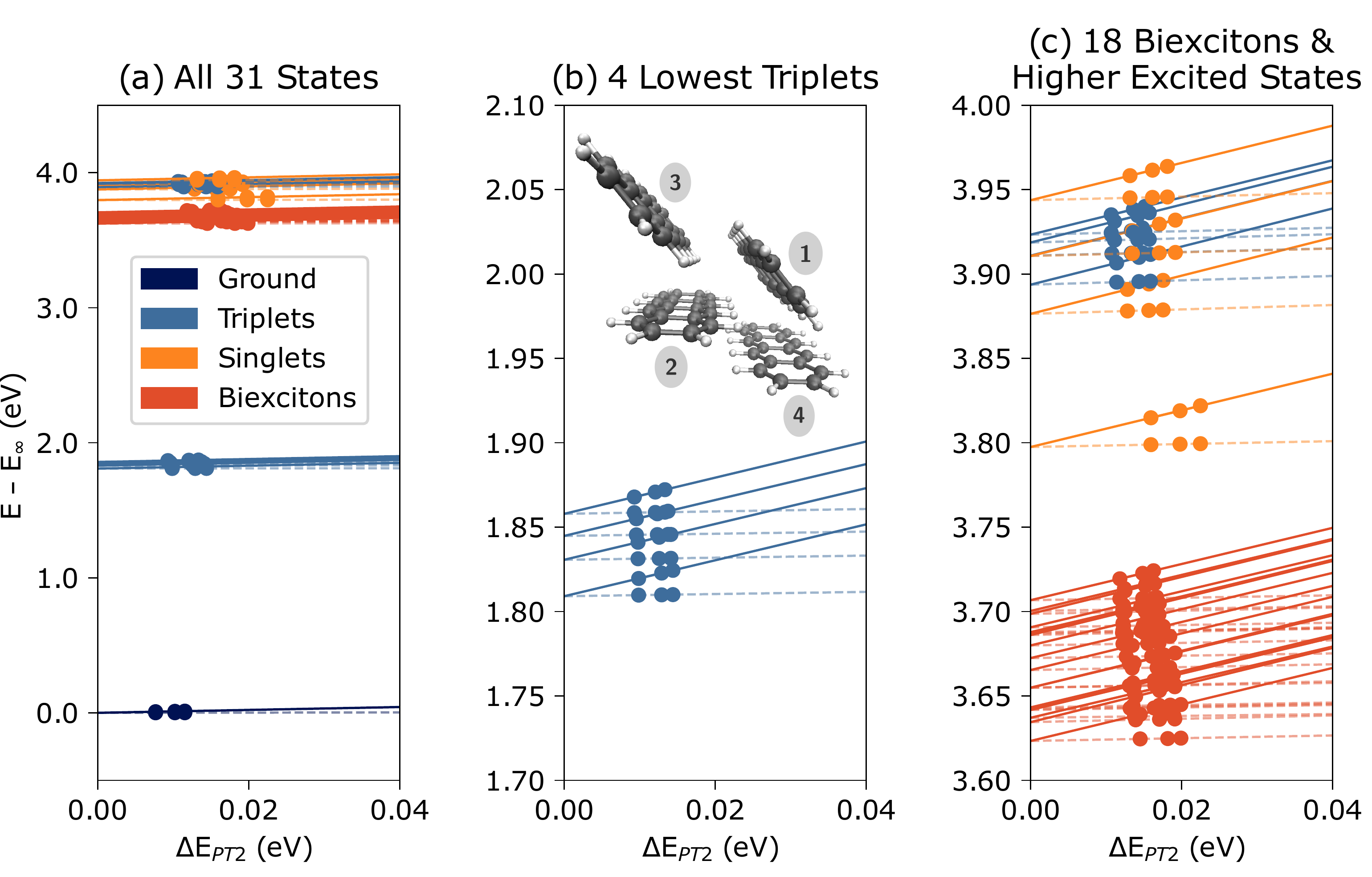}
    \caption{Extrapolated results using HOSVD bootstrapping then clip at larger thresholds to obtain extrapolation for tetracene tetramer singlet-fission example with root tracking. (a) the full spectra with 31 roots shown. (b) the middle section of the energy spectra with 4 triplet excited states and cluster labels of the tetracene tetramer. (c) the top portion of the spectra with the remaining 26 roots shown.}
    \label{fig:tt}
\end{figure*}
\pagebreak
\subsection{Singlet Fission: Tetracene Tetramer}\label{sec:sf}
Singlet fission is a multichromophoric process where a bright singlet excited state is converted into two lower energy triplets.  
The mechanism involves an entangled multiexciton singlet state, $\TTs$.\cite{Smith2013}
While the $\TTs$ state is likely the first multiexciton state to be accessed, due to spin conservation, 
it has been recently shown that the triplet and quintet  multiexcitons, $\TTt$ and $\TTq$, also play an important role in the separation process.\cite{johnsonOpenQuestionsPhotophysics2021}
Because of the intrinsic two-electron nature of the multiexcitonic state, it is difficult to compute all three spin states of the multiexciton, the initial singlet excitation, and the final triplet states on equal footing.
However, because of the underlying product structure of the target states,\cite{morrisonEvidenceSingletFission2017} tensor product state methods offer unique advantages.
Since the chromophores are naturally partitioned into different clusters, a diabatic basis can be naturally formed using the cluster states.\cite{parkerCommunicationActivespaceDecomposition2013}
Here we test our tensor product based method on tetracene tetramer, taken from a tetracene crystal that exhibits this singlet fission process.

To construct an orbital active space that accurately represents the targeted states, we performed a CIS calculation for the first four singlets and triplets, 
then built a state-averaged one particle reduced density matrix (1RDM) and diagonalized to obtain the natural orbitals.\cite{cisno}
Using the eigenvalues of the state-averaged 1RDM, the 40 most correlated orbitals (those with the most fractional occupations) are taken as our active space.
While a larger active space would have been possible in principle, 
this is the largest active space that was tractable when treating each chromophore (10 electrons in 10 orbitals) with an exact FCI cluster solver.
In future work, we will report on a restricted active space configuration interaction (RASCI) cluster solver to increase the size of the clusters (and thus active spaces) treatable. 
After defining the (40o, 40e) orbital active space, we constructed an initial guess through localization, then variationally optimized the cluster orbitals with cMF, defined by 4 clusters each with 10 electrons ($5\alpha$ + $5\beta$) in 10 orbitals.


\subsubsection{Extrapolation}
We use the same technique that was used for the PAH systems to obtain the extrapolated plots seen in Figure \ref{fig:tt}. 
In subplot (a) of Figure \ref{fig:tt}, we show 31 states that were calculated using TPSCI.
We label the states based on the expectation value of the $\hat{S}^2$ operator and dominant Fock space configurations in each eigenstate.
The singlet ground state is denoted in navy blue, triplets are in blue, singlets (bright states) are in orange, and the biexcitons are in red. 
The orientation of tetracene tetramer is shown in a herringbone lattice where the chromophores are stacked then shifted slightly from one another.
We observe faster convergence for the lower excited states compared to the biexcitons and higher single excitons which follows the intuition that the higher excited state manifold is generally more entangled, thus requiring more TPS configurations to converge. 
In subplot (b) of Figure \ref{fig:tt}, we reduce the energy scale to highlight the four lowest energy triplet states. Because these states are largely ``singly excitonic" in nature, their rate of convergence is similar to the convergence of the ground state. 

In subplot (c), the energy scale is changed to highlight the higher energy states, including all 18 biexcitons, four singlets, and four higher excited triplet states (primarily superpositions of $T2$ single excitons) are shown. 
As expected, the biexciton spectrum (shown in red) is relatively dense compared to the higher energy states.

Due to the fact that we have used a fixed orbital active space of 40 orbitals and neglected external correlation, our excitation energies are expected to be significantly overestimated compared to experiment. 
For example, the experimental value of the bright state lies at 2.35 eV.\cite{Wilson2013} 
The significant difference between the experimental value and our computed results, 
is primarily due to both the inadequate composition of our active space and missing vibronic effects, not from errors within the active space.

In the future, we would plan to use TPSCI as a CASSCF solver, allowing us to use state averaging so that our active space orbitals treat the ground and excited states on an equal footing. We note that the orbital optimization during the cMF calculation only mixes the active 40 orbitals among themselves, but this could be extended to mix all the orbitals. 
In addition to orbital optimization, we will also consider the inclusion of dynamical correlation via an operator downfolding (e.g., DUCC\cite{Bauman2019,Karol2020}) or by doing a PT2 type correction.
Finally, we can also increase the size of our active space beyond this 40 orbital example. 
This will require a more efficient cluster state solver to allow us to exceed the 10 orbitals per cluster used in this calculation. 
A RASCI cluster solver enabling larger clusters will be reported in a subsequent manuscript. 


\begin{table}[]
\begin{center}
\caption{Results for all 31 eigenstates of tetracene tetramer with associated labels based on expectation values of the $\hat{S}^2$ operator $\expval{\hat{S}^2}$, variational excitation energies ($\omega_\text{Var}$), PT2 energy corrections ($\omega_\text{PT2}$) for excitation energies in eV, and percentage of charge transfer (\% CT) for all 31 eigenstates in the TPSCI wavefunction.}
\label{tbl:wfn_allct}
\begin{tabular}{|c|c|c|c|c|c|}
\hline 
State & Label & $\expval{\hat{S}^2}$ & $\omega_\text{Var}$ & $\omega_\text{PT2}$ &  \% CT\\ \hline
1	&	S$_0$	&	0.000	&	0	&	0	&	0.09	\\
2	&	T$_1$	&	2.000	&	1.811	&	0.002	&	1.36	\\
3	&	T$_1$	&	2.000	&	1.833	&	0.002	&	0.17	\\
4	&	T$_1$	&	2.000	&	1.847	&	0.002	&	0.40	\\
5	&	T$_1$	&	2.000	&	1.860	&	0.002	&	0.36	\\
6	&	$\TTs$	&	0.001	&	3.631	&	0.007	&	3.81	\\
7	&	$\TTs$	&	0.091	&	3.642	&	0.006	&	1.95	\\
8	&	$\TTt$	&	1.919	&	3.643	&	0.006	&	1.53	\\
9	&	$\TTq$	&	5.945	&	3.648	&	0.006	&	1.36	\\
10	&	$\TTs$	&	0.234	&	3.649	&	0.006	&	4.23	\\
11	&	$\TTt$	&	1.811	&	3.650	&	0.006	&	3.02	\\
12	&	$\TTt$	&	1.838	&	3.661	&	0.005	&	3.01	\\
13	&	$\TTs$	&	0.163	&	3.661	&	0.006	&	0.98	\\
14	&	$\TTt$	&	2.000	&	3.672	&	0.006	&	0.65	\\
15	&	$\TTq$	&	5.999	&	3.678	&	0.005	&	1.21	\\
16	&	$\TTq$	&	5.998	&	3.685	&	0.005	&	0.79	\\
17	&	$\TTq$	&	5.995	&	3.691	&	0.005	&	0.52	\\
18	&	$\TTs$	&	0.302	&	3.692	&	0.005	&	0.52	\\
19	&	$\TTt$	&	1.718	&	3.693	&	0.005	&	0.50	\\
20	&	$\TTq$	&	5.986	&	3.696	&	0.005	&	0.39	\\
21	&	$\TTs$	&	0.050	&	3.704	&	0.005	&	0.96	\\
22	&	$\TTt$	&	1.961	&	3.705	&	0.005	&	0.89	\\
23	&	$\TTq$	&	5.989	&	3.711	&	0.004	&	0.42	\\
24	&	S$_1$	&	0.000	&	3.807	&	0.008	&	10.80	\\
25	&	S$_1$	&	0.000	&	3.883	&	0.005	&	1.92	\\
26	&	T$_2$	&	2.000	&	3.898	&	0.004	&	2.03	\\
27	&	T$_2$	&	2.000	&	3.915	&	0.003	&	0.36	\\
28	&	S$_1$	&	0.001	&	3.917	&	0.006	&	2.55	\\
29	&	T$_2$	&	2.000	&	3.923	&	0.003	&	0.98	\\
30	&	T$_2$	&	2.000	&	3.927	&	0.003	&	2.32	\\
31	&	S$_1$	&	0.000	&	3.950	&	0.005	&	5.03	\\ \hline
\end{tabular}
\end{center}
\end{table}

\begin{table}[]
\begin{center}
\caption{Charge transfer wavefunction analysis for first singlet excited state. These are the charge transfer Fock space configurations that contribute with a weight greater than 0.001 and are in descending order by their contributions.}
\label{tbl:wfn_ct}
\begin{tabular}{|c|c|c|c|}
\hline 
Fock Space ($\alpha$, $\beta$)  & \# Configs  & Weight & CT Character\\  \hline
( 5,5 )( 5,5 )( 5,5 )( 5,5 )	&	11157	&	0.87	&	no CT	\\
( 4,5 )( 5,5 )( 5,5 )( 6,5 )	&	1150	&	0.018	&	$1 \rightarrow 4 \hspace{2pt} (\alpha)$	\\
( 5,4 )( 5,5 )( 5,5 )( 5,6 )	&	1135	&	0.018	&	$1 \rightarrow 4 \hspace{2pt} (\beta)$	\\
( 5,4 )( 5,6 )( 5,5 )( 5,5 )	&	856	&	0.017	&	$1 \rightarrow 2 \hspace{2pt} (\beta)$	\\
( 4,5 )( 6,5 )( 5,5 )( 5,5 )	&	897	&	0.016	&	$1 \rightarrow 2 \hspace{2pt} (\alpha)$	\\
( 6,5 )( 4,5 )( 5,5 )( 5,5 )	&	843	&	0.015	&	$2 \rightarrow 1 \hspace{2pt} (\alpha)$	\\
( 5,6 )( 5,4 )( 5,5 )( 5,5 )	&	876	&	0.015	&	$2 \rightarrow 1 \hspace{2pt} (\beta)$	\\
( 6,5 )( 5,5 )( 5,5 )( 4,5 )	&	922	&	0.004	&	$4 \rightarrow 1 \hspace{2pt} (\alpha)$	\\
( 5,6 )( 5,5 )( 5,5 )( 5,4 )	&	865	&	0.004	&	$4 \rightarrow 1 \hspace{2pt} (\beta)$\\
( 5,5 )( 6,5 )( 4,5 )( 5,5 )	&	1114	&	0.004	&	$3 \rightarrow 2 \hspace{2pt} (\alpha)$	\\
( 5,5 )( 5,6 )( 5,4 )( 5,5 )	&	1099	&	0.004	&	$3 \rightarrow 2 \hspace{2pt} (\beta)$	\\
( 5,5 )( 4,5 )( 6,5 )( 5,5 )	&	951	&	0.003	&	$2 \rightarrow 3 \hspace{2pt} (\alpha)$\\
( 5,5 )( 5,4 )( 5,6 )( 5,5 )	&	938	&	0.003	&	$2 \rightarrow 3 \hspace{2pt} (\beta)$ \\ \hline
\end{tabular}
\end{center}
\end{table}
\subsubsection{Wavefunction Analysis}
To analyze the TPSCI wavefunction, we have access to the expectation value of $\hat{S}^2$, number of important configurations in each Fock space, and the associated weight of that Fock space in the overall TPS wavefunction.
In Table \ref{tbl:wfn_allct}, we present a summarized version of the wavefunction analysis.
For each state, we list the state label,  $\expval{\hat{S}^2}$, the variational excitation energies, the PT2 corrections, and the overall percent charge transfer character.

The state label is defined by the $\expval{\hat{S}^2}$ and the dominate Fock space configurations. 
While the states were generally easy to label, 
the presence of near-degeneracies between states of different spin multiplicity creates difficulties resolving spin states accurately, 
as our variational space would need to be converged to within the energy gap between the states. 
While we could always ``un-mix'' them manually by just diagonalizing the 2x2 $\hat{S}^2$ matrix, we have not investigated that in this study.


 

By analyzing the total weight of Fock sectors having clusters with different numbers of electrons, we can quantify the amount of charge transfer present in a given state. 
Following this approach, we observe a significant amount of charge transfer present in the first bright state (state 24) with 10.8 percent charge transfer.
We analyze this charge transfer character in the first singlet excited state more carefully in Table \ref{tbl:wfn_ct}, where we list charge transfer Fock space contributions that are overall greater than 0.001 in the final TPSCI wavefunction. 
As shown in this table, the localized representation of the TPSCI method makes analysis more direct. Not only can we quantify the amount of CT character, we can further decompose it into individual CT contributions. 
For example, in Table \ref{tbl:wfn_ct} we can see that while charge transfer between clusters 1 and 2 are of a ``charge resonance'' type form, where the transfers are equal in both directions, the CT interactions between clusters 1 and 4 are more asymetrical, with more electron density moving from 1 to 4 than in the opposite direction.

\section{Conclusion}
In this work, we generalize our Tensor Product Selected Configuration Interaction algorithm to enable the computation of excited states.
TPSCI has the ability to provide extremely accurate (near-exact FCI) results for strongly correlated systems that would otherwise be intractable for Slater determinant based methods.  We use our own extrapolated results within the limited basis of active orbitals as a benchmarking tool for our TPSCI method as no experimental or FCI results can be obtained for the numerous excited states computed. 

We demonstrated the accuracy of TPSCI for excited states on a series of small polycyclic aromatic hydrocarbons (PAH) molecules with active spaces of 24 electrons in 24 orbitals.
The excitation energies are within 1 kcal/mol or 0.043 eV once the state specific PT2 correction is added for the P1, P3, and P4 systems when compared to our near-exact extrapolated results.
The P2 system was a very challenging system due to the additional connectivity that increases inter-cluster interactions.
Nonetheless, all but four higher energy excited states were  within 0.043 eV of the  extrapolated results.
We see from the extrapolated results for P2, one of the higher excited states has an extreme slope  suggesting the further development of a quasidegenerate PT2 formulation.
We then extend TPSCI to one larger PAH system, P5, with an active space of 36 electrons in 36 orbitals and compare to semi-stochastic heat bath CI (SHCI).
All TPSCI excitation energies for P5 are extremely close to the near exact extrapolated result. 
When we compare TPSCI to SHCI for the P5 system, we are able to calculate an additional 12 states with TPSCI.  
Furthermore, all variational excitation energies with TPSCI are closer to their respective extrapolated results compared to how far away HCI excitation energies are from their extrapolated values.
TPSCI also has smaller PT2 corrections when compared to SHCI for all states.

After testing TPSCI on smaller PAH systems and comparing to SHCI, we investigated TPSÇI's ability to compute ``beyond-dimer model'' singlet fission excited states of a tetracene tetramer cluster. 
We chose an activate space of 40 electrons in 40 orbitals with a total of four clusters (one for each tetracene). 
Our model has all three spin states of the dark multi-exciton state as well as the singlet excited states.
We calculated the ground state and 30 excited states (eight triplets, four singlets, and 18 biexcitons).
All variational excitation energies are extremely accurate and even closer to exact results with the PT2 energy correction (only 0.001 eV difference).
 We are able to label our states from both the $\expval{\hat{S}^2}$ values and TPS wavefunction analysis. 

In addition to accurately solving large active spaces for several roots, 
the TPSCI wavefunction further allows analysis of charge-transfer character and multi-exciton states.
This analysis will be extended to produce quantitiate diabatic bases and subsequent effective Hamiltonians. 
The TPS representation also makes analysis in terms of quantum information quantities like von Neumann entropy very natural. These directions, in addition to the construction of properties and RDMs, will be the focus of future work. 

In order to extend TPSCI to larger active spaces, 
it will be necessary to use an approximate solver within the cluster, like the restricted active space approach, which will be the the focus of a future manuscript. 
In addition to improved cluster solvers, automation of the orbital clustering is also needed to minimize the amount of user input needed to setup a calculation. 
Even though we report  excitation energies near the exact limit within the active space, we have not yet included any influence from the higher lying virtual orbitals which is necessary for recovering  dynamic correlation.
Including this external dynamic correlation will be the focus of future work, either through downfolding or PT2 treatments.
Orbital optimization with the TPSCI method is also a possible future direction to provide CASSCF values for large active spaces.
Even without these suggested future directions, TPSCI has the ability to study ground states, excited states, charge-transfer states, and multiexciton states for large, strongly correlated systems and hopes to serve as an accurate method to benchmark against for systems that are intractable with FCI.

\section{Acknowledgments}
This material is based upon work supported by the National Science Foundation (Award No. 1752612).

\appendix

\section{Definition of Perturbation theory}\label{app:pt}
The perturbation theory used in this work is defined by using L\"owdin's partitioning theory.
We seek a correction to the zeroth-order wavefunction for state $s$, which is constructed as a linear combination of TPS's that lie within the $\mathcal{P}$ space. We refer to this reference state as $\ket{\mathcal{P}^s}$.
To partition the Hamiltonian for perturbative treatment,
\begin{align}
    \hat{H} = \hat{H}^{(0)} + \lambda\hat{H}^{(1)}
\end{align}
we wish to choose a partitioning where the zeroth-order contribution contains the full Hamiltonian in the $\mathcal{P}$ space, but an approximate, diagonal Hamiltonian in the $\mathcal{Q}$ space. 
This is achieved by the following partitioning:
\renewcommand{\arraystretch}{1.5}
\begin{align}
    \hat{H}^{(0)} =&\left(
    \begin{array}{c|c}
    \hat{H} & 0\\\hline
    0 & \hat{F}_D^{cMF} + \mel{\mathcal{P}^s}{\hat{V}^{cMF}}{\mathcal{P}^s}
    \end{array}\right)\\
    \hat{H}^{(1)} =&
    \left(\begin{array}{c|c}
    0 & \hat{H}\\\hline
    \hat{H} & \hat{H} - \hat{F}_D^{cMF} - \mel{\mathcal{P}^s}{\hat{V}^{cMF}}{\mathcal{P}^s}
    \end{array}\right),
    \end{align}
    where $\hat{H} = \hat{F}^{cMF}_D + \hat{V}^{cMF}$, and
    \begin{align}
    \hat{F}^{cMF} = \sum_I\hat{H}_I^{eff},
\end{align}
and where the subscript $D$ ($\hat{F}^{cMF}_D$) indicates the diagonal of the operator (this is only consequential if one is working in the HOSVD basis because the cMF effective Hamiltonian is is already diagonal in the cMF basis). 
This is referred to as a ``barycentric'' partitioning, because the zeroth-order Hamiltonian contains the reference state expectation value of the ``Fock-like'' cMF Hamiltonian. 

With this partitioning, the expression for the first order coefficients becomes:
\begin{align}
    c_j^{s(1)} = \frac{
\mel{\mathcal{Q}_j}{\hat{H}}{\mathcal{P}^s}
}{
\mel{\mathcal{P}^s}{\hat{F}^{cMF}}{\mathcal{P}^s}
-
\mel{\mathcal{Q}_j}{\hat{F}_D^{cMF}}{\mathcal{Q}_j}
}    
\end{align}
While other partitionings are likely to work better (Epstein-Nesbet\cite{epstein1926stark,nesbet1955configuration} or even a quasidegenerate formulation of the above approach\cite{mayhallIncreasingSpinflipsDecreasing2014, mayhallQuasidegenerate2ndorderPerturbation2014}), we defer consideration of different partitionings to future work. 

\section{Computing the first order wavefunction}\label{app:pt1}
The first-order correction shows up in two distinct places in the TPSCI algorithm: Step 3 and Step 5 in Sec. \ref{alg}.
In Step 5, the state specific PT2 energy correction is computed but since we only compute the final second order energy, we don't actually ever need the full first order wavefunction. Consequently, we obtain only small batches of the first order wavefunction (external configurations that share the same {\tt FockConfig}), 
then compute a batch's contribution to the energy, then immediately discard the corresponding first order amplitudes. 
In this way, computing the PT2 correction in Step 5 does not really create a memory bottleneck. 
However, in Step 3 when we perturbatively search the $\mathcal{Q}$-space to expand our $\mathcal{P}$-space, we do need to compute the first order wavefunction. 
To do this, we use prescreening to reduce the number of terms we have to consider.

To build the first order wavefunction, we apply the Hamiltonian to our varitational space, but because the Hamiltonian connects a single TPS with up to a quartic number of new TPS's ($\mathcal{O}(M^4)$), the first order interaction space quickly becomes intractable to store in memory.
For example, consider a single 4-body term in the Hamiltonian (i.e., each fermionic creation/annihilation operator is acting on a different cluster) applied to state $s$, $\ket{\Psi^s}$, is given below as:
\begin{widetext}
\begin{align}
    \ket{\sigma^s} =& \hat{H}\ket{\Psi^s} \\
    \ket{\sigma^s} \leftarrow& \hat{H}_{1,2,3,4}\ket{\Psi^s} \\
    \sigma_{\alpha'\beta'\gamma'\delta'\epsilon}^s 
    \ket{\alpha'\beta'\gamma'\delta'\epsilon}
    \leftarrow& 
    \expval{pq|rs}
    \left(
    \Gamma_{\hat{p}^\dagger}^{\alpha'\alpha}
    \dyad{\alpha'}{\alpha}
    \otimes
    \Gamma_{\hat{q}^\dagger}^{\beta'\beta}
    \dyad{\beta'}{\beta}\otimes
    \Gamma_{\hat{s}}^{\gamma'\gamma}
    \dyad{\gamma'}{\gamma}\otimes
    \Gamma_{\hat{r}}^{\delta'\delta}
    \dyad{\delta'}{\delta}
    \right)
    \ket{\alpha\beta\gamma\delta\epsilon} 
    c_{\alpha\beta\gamma\delta\epsilon}^s\\
    \sigma_{\alpha'\beta'\gamma'\delta'\epsilon}^s 
    \leftarrow& 
    \expval{pq|rs}
    \Gamma_{\hat{p}^\dagger}^{\alpha'\alpha}
    \Gamma_{\hat{q}^\dagger}^{\beta'\beta}
    \Gamma_{\hat{s}}^{\gamma'\gamma}
    \Gamma_{\hat{r}}^{\delta'\delta}
    c_{\alpha\beta\gamma\delta\epsilon}^s
    = H_{\alpha\beta\gamma\delta}^{\alpha'\beta'\gamma'\delta'}
    c_{\alpha\beta\gamma\delta\epsilon}^s
\end{align}
\end{widetext}

Because the variational states, $\ket{\Psi^s}$ are always represented in a sparse basis, this operation is performed element wise over the configurations in the variational space.
To denote this, we will rewrite the above equation, highlighting the fact that the right hand side indices are specified:
\begin{align}\label{eq:sig1}
    \sigma_{****\epsilon}^s 
    \leftarrow& 
    \sum_{pqrs}
    \expval{pq|rs}
    \Gamma_{\hat{p}^\dagger}^{*}
    \Gamma_{\hat{q}^\dagger}^{*}
    \Gamma_{\hat{s}}^{*}
    \Gamma_{\hat{r}}^{*}
    c^s
    \end{align}
    where the $*$ symbol is used to denote the full range of values of the associated index.
The above equation reveals the potential bottleneck of computing the first order wavefunction. 
Suppose $M=400$, (i.e., the number needed to keep all the states for a six orbital six electron cluster), 
 a single TPS in the variational space could couple to such a large number of configurations it would require 200Gb of memory to simply store a single
 contribution, 
$\sigma_{****\epsilon}^s$.
To avoid this, we can perform a series of screenings based on the following inequalities.\\
Assuming the following relationship:
\begin{align}
   A_{ijkl} =& B_{abcd} C_{ai} D_{bj} E_{ck} F_{dl}
\end{align}
\textbf{FOIS Screening Inequality 1}:\\
\begin{align}
   |A_{****}|_\infty \leq& \sum_{abcd} |B_{abcd}|\cdot |C_{a*}|_\infty\cdot |D_{b*}|_\infty\cdot |E_{c*}|_\infty\cdot |F_{d*}|_\infty
\end{align}
\textbf{FOIS Screening Inequality 2:}\\
\begin{align}
   |A_{***l}|_\infty \leq& \sum_{abcd} |B_{abcd}|\cdot |C_{a*}|_\infty\cdot |D_{b*}|_\infty\cdot |E_{c*}|_\infty\cdot |F_{dl}|
\end{align}
Using Inequality 1 allows us to determine if an entire block of contributions will have values all smaller than a user specified threshold $\epsilon_\text{FOIS}$, allowing us to avoid the computation altogether.
Inequality 2 allows us to predetermine which cluster state indices are capable of contributing. 
This means that after determining if the entire block is not negligible (by using Inequality 1), 
we can prune the number of cluster states, creating a smaller, effective $M$ value, so that the terms we actually compute in Eq. \ref{eq:sig1}
have a lower percentage of discarded values. 
After prescreening the indices $\alpha', \beta', \gamma',$ and $\delta'$, 
we compute the screened block of $\sigma$ contributions. 
Finally, we filter out all values with magnitudes less than $\epsilon_\text{FOIS}$,
before storing the contribution to $\sigma$ in memory.
While the screening does incur some overhead for tighter values of $\epsilon_\text{FOIS}$,
for the looser values that are often used in Step 3 (e.g. $\epsilon_\text{FOIS} = 1e-5$ or $1e-6$),
the screening can significantly speed up a calculation without having significant impact on the results. 
In future work, we will study the interplay of thresholds and performance to better understand how the 
current screening procedure works and to potentially improve over this rather straightforward approach.

\providecommand{\latin}[1]{#1}
\makeatletter
\providecommand{\doi}
  {\begingroup\let\do\@makeother\dospecials
  \catcode`\{=1 \catcode`\}=2 \doi@aux}
\providecommand{\doi@aux}[1]{\endgroup\texttt{#1}}
\makeatother
\providecommand*\mcitethebibliography{\thebibliography}
\csname @ifundefined\endcsname{endmcitethebibliography}
  {\let\endmcitethebibliography\endthebibliography}{}

\begin{figure*}
    \centering\textbf{TOC Graphic}\par\medskip
    \includegraphics[width=3.25in]{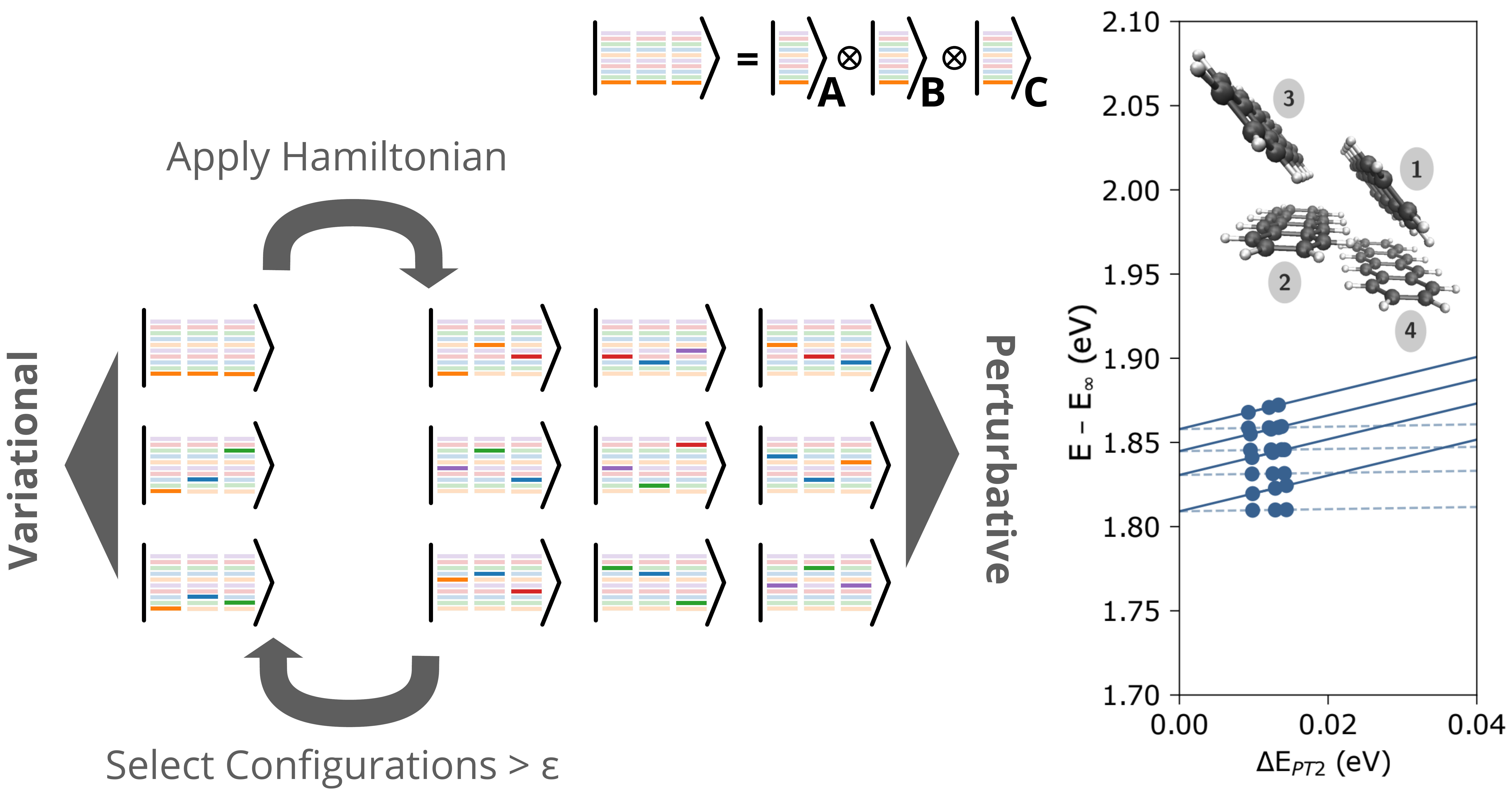}
    \label{fig:toc}
\end{figure*}
\end{document}


\beginsupplement
\title{Supporting Information: Generalization of the Tensor Product Selected CI Method for Molecular Excited States} 
\author{Nicole M. Braunscheidel}
\affiliation{Department of Chemistry, Virginia Tech,
Blacksburg, VA 24060, USA}
\author{Vibin Abraham}
\affiliation{Department of Chemistry, University of Michigan, Ann Arbor, MI 48109, USA}
\author{Nicholas J. Mayhall}
\email{nmayhall@vt.edu}
\affiliation{Department of Chemistry, Virginia Tech,
Blacksburg, VA 24060, USA}

\maketitle

\section{Comparison with TPS-SE}
Before investigating the TPSCI approach, we evaluate the
    accuracy of a cheaper method to obtain excited states, the TPS-single exciton or TPS-SE approach. 
For a TPS-SE calculation, we take a linear combination of tensor product states where one cluster has a triplet exciton as shown below. 
\begin{equation}\label{eq:tps_se}
    \ket{\psi_{t}} = \ket{t,0,0,0} + \ket{0,t,0,0} + \ket{0,0,t,0} + \ket{0,0,0,t}
\end{equation}
We analyze the accuracy of this TPS-SE method for the P1 system.
We also investigate the effect of a perturbative correction based on barycentric MP2 on top of the TPS-SE, named TPS-SE(2).
We compare the excitation energies computed using TPS-SE and TPS-SE(2) with the extrapolated TPSCI results for P1 (Table \ref{tbl:tpscis}).
As seen from Table \ref{tbl:tpscis}, neither TPS-SE or the  TPS-SE(2) method provide accurate results when compared to the near exact excitation energies ($\omega_\infty$). 
This is not surprising since there are no inter-cluster charge transfer contributions in the TPS-SE approach.
Additionally, there are only perturbatively inter-cluster contributions in TPS-SE(2).
Being a perturbative correction, the  TPS-SE(2) can also have issues when there are degenerate states similar to traditional MRPT approaches.
Meanwhile, the TPSCI approach includes higher excited configurations absent from TPS-SE(2) and does not have these degeneracy issues. 
TPSCI generates a variational space that captures a majority of the important configurations. 
As shown here, it is very important to go beyond TPS-SE.

\begin{table}[h!]
 \caption{Excitation energies (kcal/mol) for the 4 triplet excited states for the P1 system computed using CIS, TPS-SE, TPS-SE(2), TPSCI, TPSCI with PT2 corrections, and extrapolated TPSCI.}
 \label{tbl:tpscis}
 \begin{tabular}{l|ccccccc}
 \hline\hline
  & CIS & TPS-SE & TPS-SE(2) & TPSCI & TPSCI+PT2 & $\omega_\infty$ \\ \hline 
T$_1$	&	68.05	&	97.23	&	91.13	&	74.82	&	74.17	&	74.05	\\
T$_2$	&	75.15	&	97.46	&	93.09	&	81.03	&	80.46	&	80.35	\\
T$_3$	&	85.44	&	100.84	&	93.88	&	89.89	&	89.30	&	89.18	\\
T$_4$	&	94.96	&	103.15	&	100.62	&	101.01	&	100.68	&	100.59	\\
 \hline \hline
 \end{tabular}
 \end{table}

\pagebreak 
\section{TPSCI vs TPSCI with HOSVD Decomposition}

\begin{figure*}[h!]
    \includegraphics[width=1.0\linewidth]{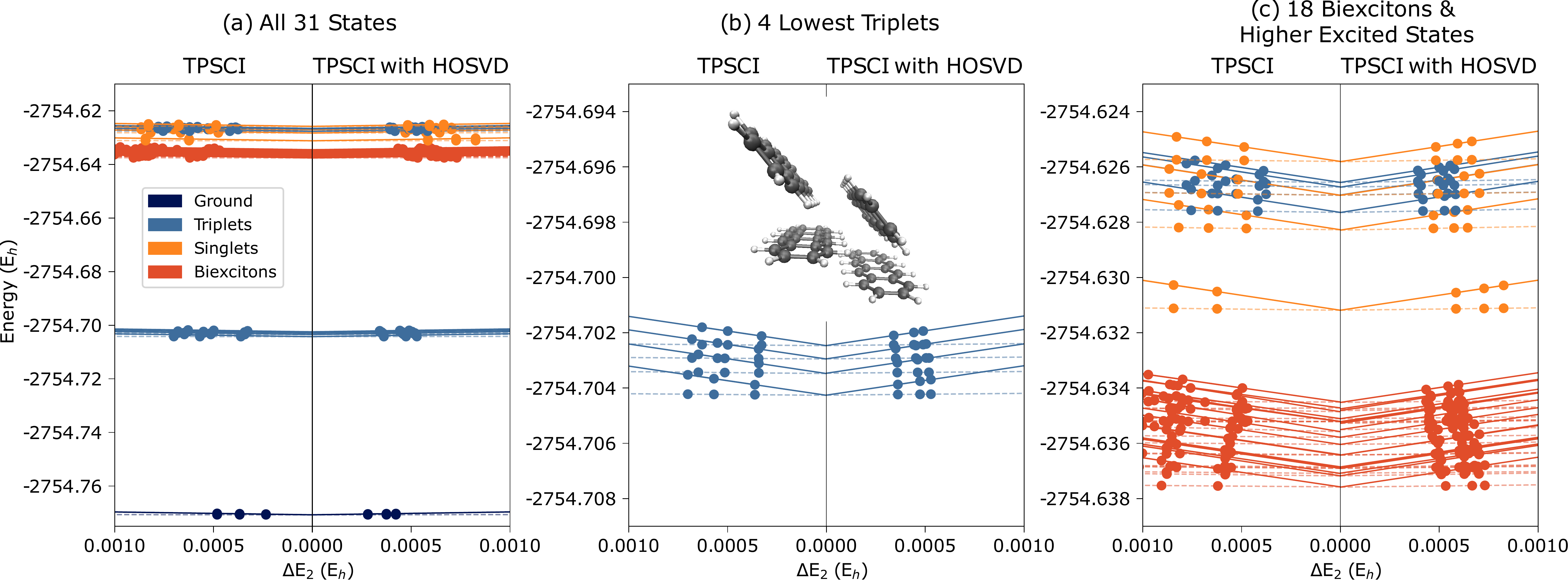}
    \caption{Extrapolated results to compare TPSCI with and without the HOSVD with bootstrapping for tetracene tetramer singlet fission example with root tracking. (a) the full spectra with 31 roots shown. (b) the middle section of the energy spectra with 4 triplet excited states. (c) the top portion of the spectra with the remaining 26 roots shown.}
    \label{fig:compare_tt}
\end{figure*}

The decision to use the HOSVD rotated basis then clip the remaining coefficients in the wavefunction to obtain the extrapolated results came from the following analysis of TPSCI (from cMF reference) verses TPSCI with the HOSVD basis. 
We show results only for the tetracene tetramer but similar result were found for the PAH systems.
In Figure \ref{fig:compare_tt}, we plot the extrapolated results for tetracene tetramer with TPSCI (no HOSVD) on the left and TPSCI (with HOSVD) on the right.
We can see that TPSCI with HOSVD converges faster than without HOSVD.
This is very obvious in the first singlet excited state in orange around -2754.63 Hartrees in subplot (c) of Figure \ref{fig:compare_tt}.
Additionally, the variational dimension at the tightest threshold of $\epsilon_{CIPSI} = 4\times 10^{-4}$ for TPSCI without HOSVD was 72,728 while the dimension with the HOSVD was only 35,461. 

\pagebreak
\section{Convergence of Cluster Basis Size}
\begin{figure}[h!]
    \includegraphics[width=0.5\textwidth]{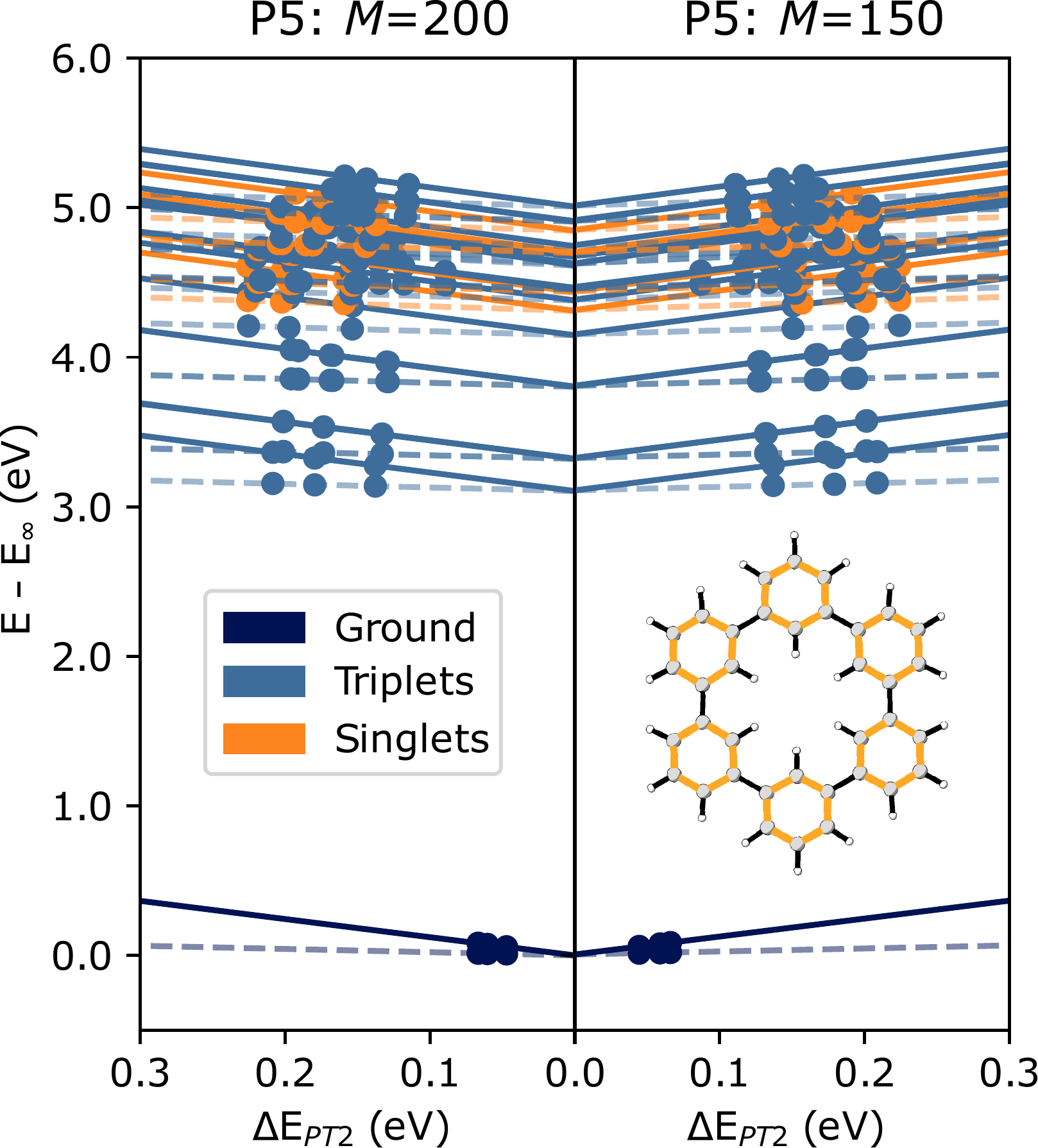}
    \caption{Comparison of extrapolated excitation energies to compare convergence of cluster basis size using TPSCI. On the left are the results for $M=200$ and on the right $M=150$. }
    \label{fig:m200vs150}
\end{figure}

The maximum number of roots ($M$) per cluster per Fock sector determines the size of the cluster eigenbasis that is accessible by the selected CI algorithm.  If this value is set too low such that the full Hilbert space is not represented, then the extrapolated results can't be considered near exact or FCI results.  To ensure that our parameter of $M=150$ was sufficient, we compared the results of various TPSCI extrapolations with $M=200, 300, $ and $400$.  We saw no significant difference in the extrapolated excitation energies by adding more roots greater than $M=200$. Additionally, when we compared $M=150$ to $M=200$, we also did not see a significant difference in the extrapolated excitation energies, slope of the extrapolations, or dimension of the TPSCI wavefunction. This can be seen in Figure \ref{fig:m200vs150}. Thus concluding that for the P5 system, a parameter of $M=150$ is sufficient to represent the full Hilbert space.

\pagebreak

\section{PAH systems}
\begin{figure*}[h!]
    \includegraphics[width=0.95\linewidth]{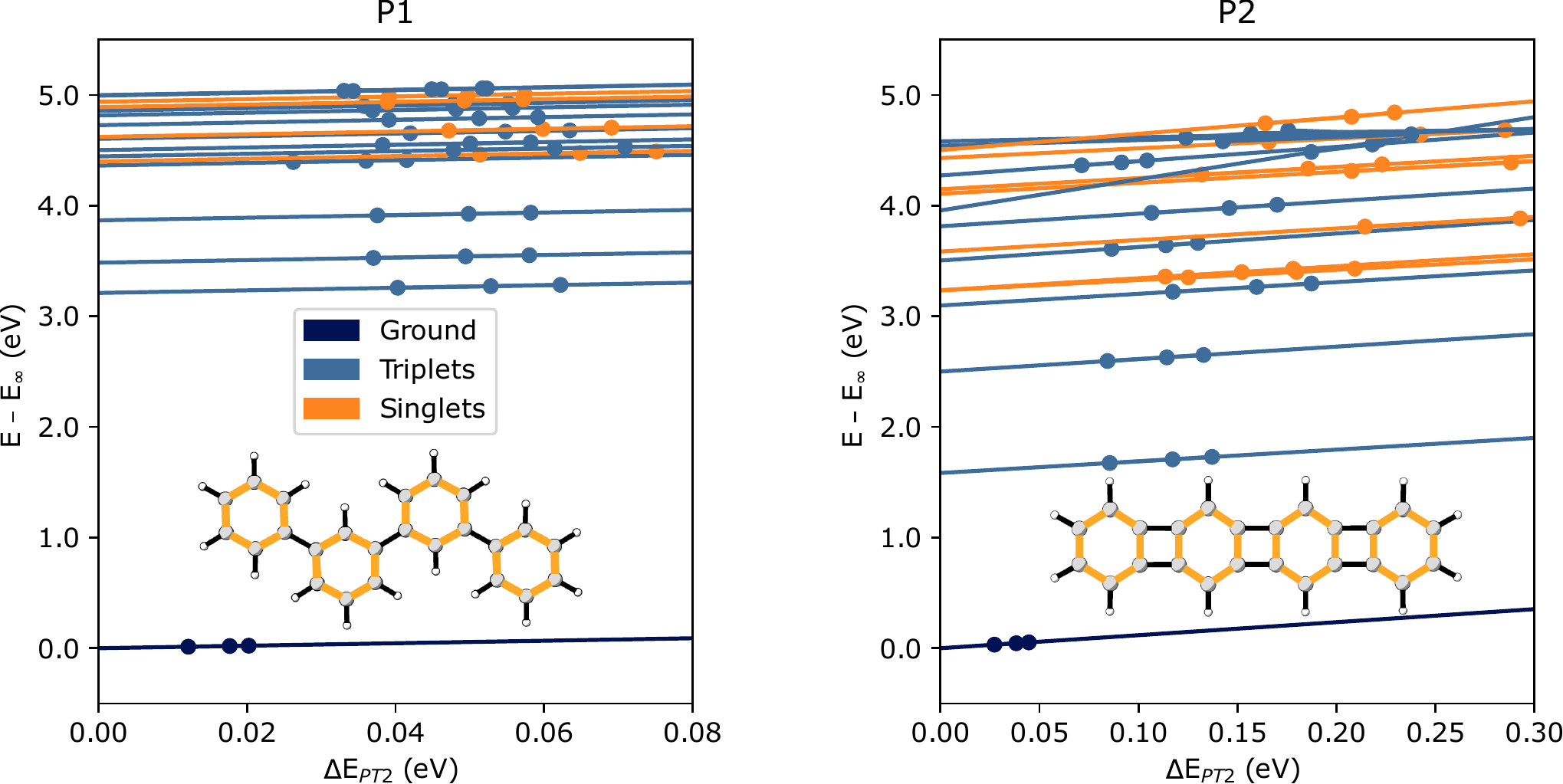}
    \caption{Variational linear extrapolations of the ground state and 16 excited states for the P1 and P2 PAH systems studied using TPSCI (with HOSVD bootstrapping). All energies are shifted by the extrapolated ground state TPSCI energy so the ground state converges to 0 eV. Each point represents a converged TPSCI calculation at a series of $\epsilon_\text{CIPSI} = n \times 10^{-4}$ with n=4,6,8 for P1 and $n \times 10^{-4}$ with n=6,8,10 for P2. Note the x-scale of the P1 system being scaled down to show the slope of the extrapolations.}
\end{figure*}

\begin{figure*}[h!]
    \includegraphics[width=0.95\linewidth]{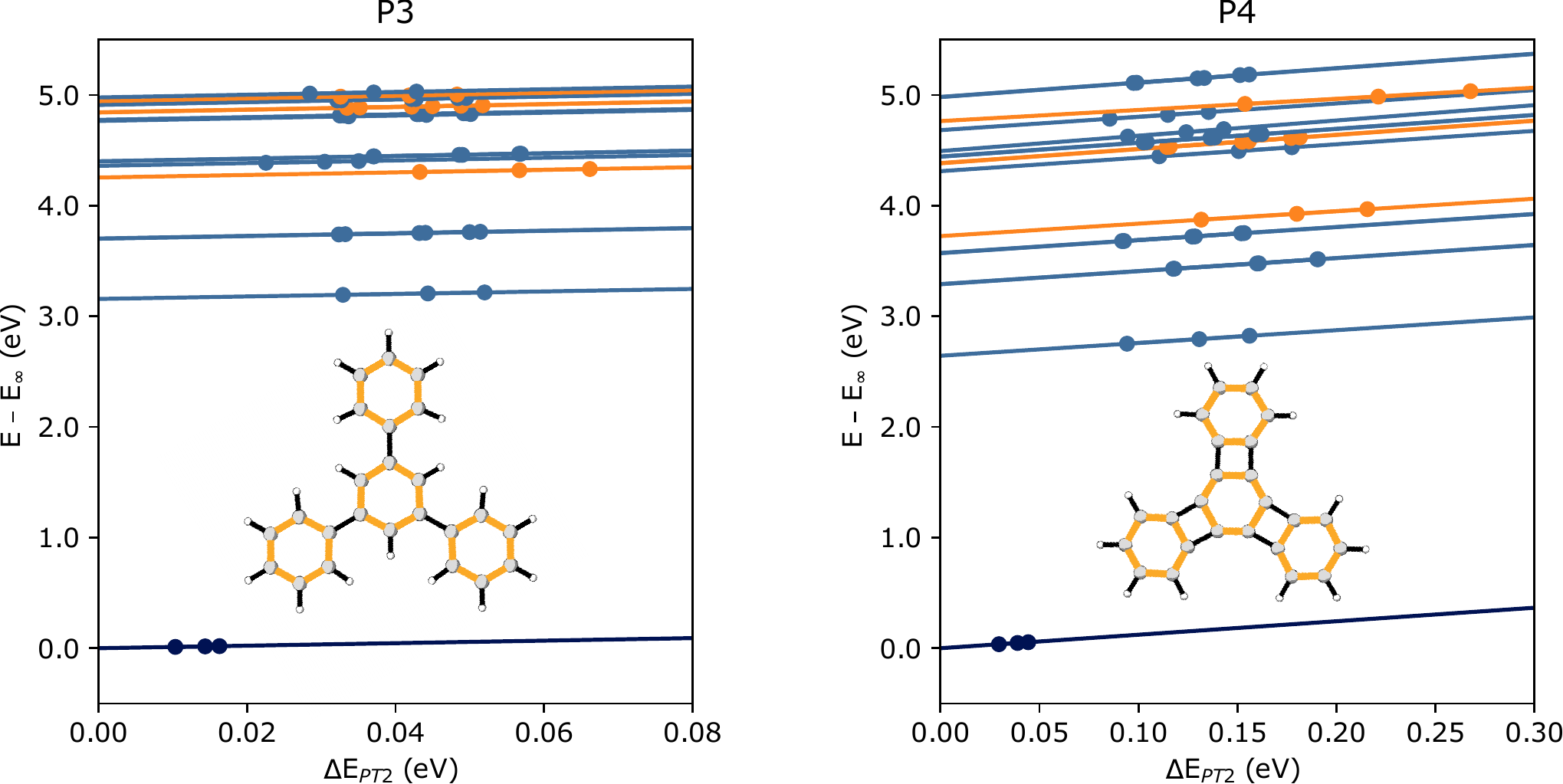}
    \caption{Variational linear extrapolations of the ground state and 16 excited states for the P3 and P4 PAH systems studied using TPSCI (with HOSVD bootstrapping). All energies are
shifted by the extrapolated ground state TPSCI energy so the ground state converges to 0 eV. Each point represents a converged TPSCI calculation at a series of $\epsilon_\text{CIPSI} = n \times 10^{-4}$ with n=4,6,8 for P3 and P4. Note the x-scale of the P3 system being scaled down to show the slope of the extrapolations.}
\end{figure*}

\pagebreak
\section{Hamiltonian Terms}
\begin{table}
\begin{center}
\caption{Enumeration of all the distinct terms for a given pair, triple, and quadruple of clusters, respectively. 
Here, $\alpha$ ($\alpha^\dagger$) refers to  annihilation (creation) of an $\alpha$ electron, and $\beta$ ($\beta^\dagger$) refers to a $\beta$ annihilation (creation) operator.}
\label{sitbl:hamterms}
\begin{tabular}{|c|p{0.3cm} p{1.5cm} p{1.5cm}|p{0.3cm} p{1cm} p{1cm} p{1cm}|p{0.3cm} p{0.75cm} p{0.75cm} p{0.75cm} p{0.75cm}|}
\hline 

Term & \multicolumn{3}{c|}{2 Body Terms}  & \multicolumn{4}{c|}{3 Body Terms} & \multicolumn{5}{c|}{4 Body Terms}\\ \hline
1	&		&	$\alpha$	&	$\alpha^{\dagger}$	&		&	$\alpha^{\dagger}$	&	$\alpha\beta$	&	$\beta^{\dagger}$	&		&	$\beta^{\dagger}$	&	$\beta$	&	$\alpha$	&	$\alpha^{\dagger}$	\\
2	&		&	$\alpha^{\dagger}\alpha\alpha$	&	$\alpha^{\dagger}$	&		&	$\beta$	&	$\beta^{\dagger}\beta^{\dagger}$	&	$\beta$	&		&	$\alpha$	&	$\alpha^{\dagger}$	&	$\beta$	&	$\beta^{\dagger}$	\\
3	&		&	$\alpha$	&	$\alpha^{\dagger}\alpha^{\dagger}\alpha$	&		&	$\alpha$	&	$\beta^{\dagger}\beta$	&	$\alpha^{\dagger}$	&		&	$\beta^{\dagger}$	&	$\beta^{\dagger}$	&	$\beta$	&	$\beta$	\\
4	&		&	$\alpha$	&	$\alpha^{\dagger}\beta^{\dagger}\beta$	&		&	$\alpha$	&	$\alpha^{\dagger}\alpha$	&	$\alpha^{\dagger}$	&		&	$\alpha^{\dagger}$	&	$\beta^{\dagger}$	&	$\alpha$	&	$\beta$	\\
5	&		&	$\beta^{\dagger}\beta\alpha$	&	$\alpha^{\dagger}$	&		&	$\beta$	&	$\beta$	&	$\beta^{\dagger}\beta^{\dagger}$	&		&	$\beta^{\dagger}$	&	$\alpha$	&	$\alpha^{\dagger}$	&	$\beta$	\\
6	&		&	$\beta\beta$	&	$\beta^{\dagger}\beta^{\dagger}$	&		&	$\beta^{\dagger}$	&	$\alpha\beta$	&	$\alpha^{\dagger}$	&		&	$\alpha^{\dagger}$	&	$\alpha$	&	$\beta$	&	$\beta^{\dagger}$	\\
7	&		&	$\beta^{\dagger}$	&	$\beta$	&		&	$\alpha^{\dagger}\alpha$	&	$\beta^{\dagger}$	&	$\beta$	&		&	$\alpha$	&	$\beta^{\dagger}$	&	$\alpha^{\dagger}$	&	$\beta$	\\
8	&		&	$\beta^{\dagger}$	&	$\alpha^{\dagger}\beta\alpha$	&		&	$\beta^{\dagger}\beta$	&	$\beta^{\dagger}$	&	$\beta$	&		&	$\beta$	&	$\beta$	&	$\beta^{\dagger}$	&	$\beta^{\dagger}$	\\
9	&		&	$\beta^{\dagger}$	&	$\beta^{\dagger}\beta\beta$	&		&	$\beta^{\dagger}\alpha$	&	$\alpha^{\dagger}$	&	$\beta$	&		&	$\beta^{\dagger}$	&	$\alpha^{\dagger}$	&	$\beta$	&	$\alpha$	\\
10	&		&	$\beta^{\dagger}\beta^{\dagger}\beta$	&	$\beta$	&		&	$\beta$	&	$\beta^{\dagger}\alpha$	&	$\alpha^{\dagger}$	&		&	$\alpha^{\dagger}$	&	$\alpha^{\dagger}$	&	$\alpha$	&	$\alpha$	\\
11	&		&	$\alpha^{\dagger}\beta^{\dagger}\alpha$	&	$\beta$	&		&	$\alpha^{\dagger}$	&	$\beta^{\dagger}$	&	$\alpha\beta$	&		&	$\alpha$	&	$\alpha^{\dagger}$	&	$\beta^{\dagger}$	&	$\beta$	\\
12	&		&	$\alpha\beta$	&	$\alpha^{\dagger}\beta^{\dagger}$	&		&	$\alpha$	&	$\alpha^{\dagger}$	&	$\alpha^{\dagger}\alpha$	&		&	$\alpha$	&	$\alpha^{\dagger}$	&	$\alpha^{\dagger}$	&	$\alpha$	\\
13	&		&	$\alpha^{\dagger}\beta^{\dagger}$	&	$\alpha\beta$	&		&	$\alpha$	&	$\alpha^{\dagger}$	&	$\beta^{\dagger}\beta$	&		&	$\beta$	&	$\beta^{\dagger}$	&	$\alpha$	&	$\alpha^{\dagger}$	\\
14	&		&	$\alpha^{\dagger}\alpha^{\dagger}$	&	$\alpha\alpha$	&		&	$\beta^{\dagger}\beta$	&	$\beta$	&	$\beta^{\dagger}$	&		&	$\alpha^{\dagger}$	&	$\beta^{\dagger}$	&	$\beta$	&	$\alpha$	\\
15	&		&	$\beta^{\dagger}\beta$	&	$\alpha^{\dagger}\alpha$	&		&	$\alpha^{\dagger}\alpha$	&	$\beta$	&	$\beta^{\dagger}$	&		&	$\alpha$	&	$\beta$	&	$\beta^{\dagger}$	&	$\alpha^{\dagger}$	\\
16	&		&	$\beta^{\dagger}\beta$	&	$\beta^{\dagger}\beta$	&		&	$\alpha^{\dagger}$	&	$\beta$	&	$\beta^{\dagger}\alpha$	&		&	$\alpha^{\dagger}$	&	$\alpha$	&	$\beta^{\dagger}$	&	$\beta$	\\
17	&		&	$\alpha^{\dagger}\alpha$	&	$\beta^{\dagger}\beta$	&		&	$\beta\beta$	&	$\beta^{\dagger}$	&	$\beta^{\dagger}$	&		&	$\beta^{\dagger}$	&	$\beta$	&	$\beta$	&	$\beta^{\dagger}$	\\
18	&		&	$\alpha^{\dagger}\alpha$	&	$\alpha^{\dagger}\alpha$	&		&	$\beta^{\dagger}$	&	$\alpha$	&	$\alpha^{\dagger}\beta$	&		&	$\alpha^{\dagger}$	&	$\alpha$	&	$\alpha^{\dagger}$	&	$\alpha$	\\
19	&		&	$\alpha^{\dagger}\alpha^{\dagger}\alpha$	&	$\alpha$	&		&	$\alpha^{\dagger}\alpha$	&	$\alpha$	&	$\alpha^{\dagger}$	&		&	$\alpha^{\dagger}$	&	$\beta$	&	$\alpha$	&	$\beta^{\dagger}$	\\
20	&		&	$\alpha^{\dagger}$	&	$\beta^{\dagger}\beta\alpha$	&		&	$\beta^{\dagger}\beta$	&	$\alpha$	&	$\alpha^{\dagger}$	&		&	$\alpha$	&	$\beta$	&	$\alpha^{\dagger}$	&	$\beta^{\dagger}$	\\
21	&		&	$\alpha^{\dagger}$	&	$\alpha^{\dagger}\alpha\alpha$	&		&	$\alpha^{\dagger}\alpha^{\dagger}$	&	$\alpha$	&	$\alpha$	&		&	$\beta$	&	$\alpha^{\dagger}$	&	$\alpha$	&	$\beta^{\dagger}$	\\
22	&		&	$\alpha^{\dagger}\beta^{\dagger}\beta$	&	$\alpha$	&		&	$\alpha\beta$	&	$\alpha^{\dagger}$	&	$\beta^{\dagger}$	&		&	$\beta$	&	$\alpha$	&	$\beta^{\dagger}$	&	$\alpha^{\dagger}$	\\
23	&		&	$\alpha^{\dagger}$	&	$\alpha$	&		&	$\beta^{\dagger}$	&	$\alpha^{\dagger}\beta$	&	$\alpha$	&		&	$\beta^{\dagger}$	&	$\beta$	&	$\beta^{\dagger}$	&	$\beta$	\\
24	&		&	$\beta^{\dagger}\beta^{\dagger}$	&	$\beta\beta$	&		&	$\alpha$	&	$\alpha^{\dagger}\alpha^{\dagger}$	&	$\alpha$	&		&	$\beta^{\dagger}$	&	$\beta$	&	$\alpha^{\dagger}$	&	$\alpha$	\\
25	&		&	$\beta$	&	$\beta^{\dagger}\beta^{\dagger}\beta$	&		&	$\alpha$	&	$\beta^{\dagger}$	&	$\alpha^{\dagger}\beta$	&		&	$\beta$	&	$\beta^{\dagger}$	&	$\beta$	&	$\beta^{\dagger}$	\\
26	&		&	$\beta^{\dagger}\beta\beta$	&	$\beta^{\dagger}$	&		&	$\alpha$	&	$\alpha$	&	$\alpha^{\dagger}\alpha^{\dagger}$	&		&	$\alpha$	&	$\alpha^{\dagger}$	&	$\alpha$	&	$\alpha^{\dagger}$	\\
27	&		&	$\beta$	&	$\beta^{\dagger}$	&		&	$\alpha^{\dagger}$	&	$\alpha^{\dagger}$	&	$\alpha\alpha$	&		&	$\beta$	&	$\alpha$	&	$\alpha^{\dagger}$	&	$\beta^{\dagger}$	\\
28	&		&	$\alpha^{\dagger}\beta\alpha$	&	$\beta^{\dagger}$	&		&	$\beta^{\dagger}$	&	$\beta$	&	$\beta^{\dagger}\beta$	&		&	$\beta^{\dagger}$	&	$\alpha$	&	$\beta$	&	$\alpha^{\dagger}$	\\
29	&		&	$\beta$	&	$\alpha^{\dagger}\beta^{\dagger}\alpha$	&		&	$\beta^{\dagger}$	&	$\beta$	&	$\alpha^{\dagger}\alpha$	&		&	$\alpha^{\dagger}$	&	$\alpha$	&	$\alpha$	&	$\alpha^{\dagger}$	\\
30	&		&	$\alpha^{\dagger}\beta$	&	$\beta^{\dagger}\alpha$	&		&	$\alpha^{\dagger}\beta^{\dagger}$	&	$\alpha$	&	$\beta$	&		&	$\alpha$	&	$\beta^{\dagger}$	&	$\beta$	&	$\alpha^{\dagger}$	\\
31	&		&	$\beta^{\dagger}\alpha$	&	$\alpha^{\dagger}\beta$	&		&	$\beta$	&	$\alpha^{\dagger}\beta^{\dagger}$	&	$\alpha$	&		&	$\alpha$	&	$\alpha$	&	$\alpha^{\dagger}$	&	$\alpha^{\dagger}$	\\
32	&		&	$\alpha\alpha$	&	$\alpha^{\dagger}\alpha^{\dagger}$	&		&	$\alpha$	&	$\alpha^{\dagger}\beta^{\dagger}$	&	$\beta$	&		&	$\alpha^{\dagger}$	&	$\beta$	&	$\beta^{\dagger}$	&	$\alpha$	\\
33	&		&		&		&		&	$\beta$	&	$\alpha$	&	$\alpha^{\dagger}\beta^{\dagger}$	&		&	$\beta$	&	$\beta^{\dagger}$	&	$\beta^{\dagger}$	&	$\beta$	\\
34	&		&		&		&		&	$\alpha^{\dagger}\beta^{\dagger}$	&	$\beta$	&	$\alpha$	&		&	$\beta$	&	$\beta^{\dagger}$	&	$\alpha^{\dagger}$	&	$\alpha$	\\
35	&		&		&		&		&	$\alpha$	&	$\beta$	&	$\alpha^{\dagger}\beta^{\dagger}$	&		&	$\beta^{\dagger}$	&	$\alpha^{\dagger}$	&	$\alpha$	&	$\beta$	\\
36	&		&		&		&		&	$\alpha^{\dagger}$	&	$\alpha$	&	$\beta^{\dagger}\beta$	&		&	$\beta$	&	$\alpha^{\dagger}$	&	$\beta^{\dagger}$	&	$\alpha$	\\
37	&		&		&		&		&	$\alpha^{\dagger}$	&	$\alpha$	&	$\alpha^{\dagger}\alpha$	&		&		&		&		&		\\
38	&		&		&		&		&	$\beta$	&	$\beta^{\dagger}$	&	$\alpha^{\dagger}\alpha$	&		&		&		&		&		\\
39	&		&		&		&		&	$\beta$	&	$\beta^{\dagger}$	&	$\beta^{\dagger}\beta$	&		&		&		&		&		\\
40	&		&		&		&		&	$\alpha^{\dagger}\beta$	&	$\beta^{\dagger}$	&	$\alpha$	&		&		&		&		&		\\
41	&		&		&		&		&	$\alpha$	&	$\alpha^{\dagger}\beta$	&	$\beta^{\dagger}$	&		&		&		&		&		\\
42	&		&		&		&		&	$\beta^{\dagger}$	&	$\alpha^{\dagger}$	&	$\alpha\beta$	&		&		&		&		&		\\
43	&		&		&		&		&	$\alpha^{\dagger}\alpha$	&	$\alpha^{\dagger}$	&	$\alpha$	&		&		&		&		&		\\
44	&		&		&		&		&	$\beta^{\dagger}\beta$	&	$\alpha^{\dagger}$	&	$\alpha$	&		&		&		&		&		\\
45	&		&		&		&		&	$\alpha^{\dagger}$	&	$\alpha\alpha$	&	$\alpha^{\dagger}$	&		&		&		&		&		\\
46	&		&		&		&		&	$\beta^{\dagger}$	&	$\alpha^{\dagger}\alpha$	&	$\beta$	&		&		&		&		&		\\
47	&		&		&		&		&	$\beta^{\dagger}$	&	$\beta^{\dagger}\beta$	&	$\beta$	&		&		&		&		&		\\
48	&		&		&		&		&	$\beta^{\dagger}$	&	$\beta^{\dagger}$	&	$\beta\beta$	&		&		&		&		&		\\
49	&		&		&		&		&	$\alpha^{\dagger}\beta$	&	$\alpha$	&	$\beta^{\dagger}$	&		&		&		&		&		\\
50	&		&		&		&		&	$\beta^{\dagger}\alpha$	&	$\beta$	&	$\alpha^{\dagger}$	&		&		&		&		&		\\
51	&		&		&		&		&	$\alpha^{\dagger}$	&	$\alpha^{\dagger}\alpha$	&	$\alpha$	&		&		&		&		&		\\
52	&		&		&		&		&	$\alpha^{\dagger}$	&	$\beta^{\dagger}\beta$	&	$\alpha$	&		&		&		&		&		\\
53	&		&		&		&		&	$\alpha\beta$	&	$\beta^{\dagger}$	&	$\alpha^{\dagger}$	&		&		&		&		&		\\
54	&		&		&		&		&	$\alpha^{\dagger}$	&	$\beta^{\dagger}\alpha$	&	$\beta$	&		&		&		&		&		\\
55	&		&		&		&		&	$\beta$	&	$\alpha^{\dagger}$	&	$\beta^{\dagger}\alpha$	&		&		&		&		&		\\
56	&		&		&		&		&	$\alpha\alpha$	&	$\alpha^{\dagger}$	&	$\alpha^{\dagger}$	&		&		&		&		&		\\
57	&		&		&		&		&	$\beta$	&	$\beta^{\dagger}\beta$	&	$\beta^{\dagger}$	&		&		&		&		&		\\
58	&		&		&		&		&	$\beta$	&	$\alpha^{\dagger}\alpha$	&	$\beta^{\dagger}$	&		&		&		&		&		\\
59	&		&		&		&		&	$\beta^{\dagger}\beta^{\dagger}$	&	$\beta$	&	$\beta$	&		&		&		&		&		\\
60	&		&		&		&		&	$\beta^{\dagger}$	&	$\beta\beta$	&	$\beta^{\dagger}$	&		&		&		&		&		\\
 \hline
\end{tabular}
\end{center}
\end{table}